\documentclass[pre,showpacs,twocolumn,amsmath,amssymb,floatfix,superscriptaddress]{revtex4}
\usepackage{graphicx}
\usepackage{dcolumn}

\usepackage{bm}
\usepackage{hyperref}
\usepackage{latexsym}

\begin{document}
\title{Voting and Catalytic Processes with Inhomogeneities}

\author{Mauro Mobilia and Ivan T. Georgiev}
\email{mmobilia@vt.edu,georgiev@vt.edu}
\affiliation{Center for Stochastic Processes in Science and Engineering, Department of Physics, 
Virginia Polytechnic Institute and State University,  Blacksburg, VA, 24061-0435, USA}
\date{\today}

\begin{abstract}

We consider the dynamics of the voter model and of the monomer-monomer
catalytic process in the presence of many ``competing'' inhomogeneities and
show, through exact calculations and numerical simulations, that their presence results in
a nontrivial
fluctuating steady state whose properties are studied and turn out to specifically depend 
on the dimensionality of the system, the strength of the inhomogeneities and their
separating distances.
In fact, in arbitrary dimensions, we obtain an exact (yet formal) expression of
the order parameters (magnetization and concentration of adsorbed particles)
in the presence of an arbitrary number $n$ of inhomogeneities (``zealots'' in the voter language) 
and formal similarities with {\it suitable electrostatic systems} are pointed out. 
In the nontrivial cases $n=1, 2$, we
explicitly compute the static and long-time properties of the order parameters and therefore capture
the generic features of the systems. When $n>2$, the problems are studied through numerical
simulations. In one spatial dimension, we also compute the expressions of the stationary order parameters in the
completely disordered case, where $n$ is arbitrary large. Particular attention is paid to 
the spatial dependence of the stationary order parameters and formal connections with electrostatics.
\end{abstract}

\maketitle

\section{Introduction}

Recently, much attention has been devoted to the field of nonequilibrium
many-body stochastic processes \cite{Privman}. In particular the study
of exact solutions of prototypical models such as the {\it voter model} \cite{Liggett} has
proved to be fruitful
for understanding a broad class of nonequilibrium phenomena \cite{Privman}.
In modeling nonequilibrium systems, it is often assumed that the underlying spatial
structure is homogeneous. However, in real situations stochastic processes take place in the presence
of imperfections (dislocations, defects, etc) that modify locally the interactions (see e.g.
\cite{Privman,dis}  and references  therein).
It is therefore highly desirable to take into account the effects of disorder,
inhomogeneities and defects or other spatial constraints within simple and mathematically amenable models.
Motivated by the above considerations, in a recent letter \cite{IVM}, the
properties of a paradigmatic nonequilibrium statistical mechanics model (the voter model) in the presence
of one single inhomogeneity (a zealot) have been studied and it was shown that the presence of single zealot has
dramatic effects on the dynamics and the steady state. 
For this model, in low dimensions, all of the agents eventually follow the zealot.
Obviously, real systems are quite complex and the case of a single defect cannot
be considered as being generic. To gain some insight on more realistic situations,
we present here an approach allowing us to compute exact properties, in arbitrary
dimensions, of a class of stochastic many-body systems in the presence of $n$ {\it competing inhomogeneities}.
This study is carried out in the context of two physically relevant systems: 
the voter model and the monomer-monomer catalytic reaction (in the reaction-controlled
limit). 
We consider such a study as a further contribution toward the understanding of a class of
disordered nonequilibrium many-body processes (where inhomogeneities would not be
spatially fixed but 
would be randomly distributed). We show that the presence of ``competing inhomogeneities''
(in the sense
of locally {\it perturbing} the otherwise homogeneous dynamics) generally results in a
space-dependent fluctuating steady state.
The amenable case where $n=2$ is analytically studied in detail and 
the static and long-time properties of the order
parameters  are obtained and their spatial dependence are computed.
The situation 
where $n\geq 2$ is investigated by numerical simulations. Also, in one spatial dimension,
we are able 
to compute  the stationary order parameters in the completely disordered case ({\it i.e}
when $n$ is arbitrary large).
We therefore show how the stationary magnetization/concentration depends on the
dimensionality of the system, 
the strength of the inhomogeneities and their separating distances.
In particular, we show that the local perturbation of the dynamics may give rise to 
subtle coarsening phenomena. In 1D and 2D, when the density of the inhomogeneities is vanishing 
in the thermodynamic limit there is still coarsening in the system. Oppositely, when the density of the competing inhomogeneities is non-zero there is no coarsening, even in one and two dimensions.
We obtain an exact, yet formal, expression of the order parameters (magnetization 
and concentration of adsorbed particles) in arbitrary dimension.
In dimensions $d=2, 3$  we pay special attention to the radial and polar dependence of
these quantities. 
Also, formal similarities with {\it electrostatic systems} are pointed out.
The organization of this work is the following: In the next section we introduce the
inhomogeneous voter model. 
In Section III, we present the general mathematical set-up and the formal solution of the
problem. In Section IV, 
we study analytically the voter model in the presence of two ``competing zealots'' in one,
two and three dimensions 
and provide numerical results for the case where $n\geq 2$. In Section IV.B, for the
one-dimensional case, 
we also derive the expression of the static magnetization in the completely disordered
situation where $n$ 
is arbitrary large.
Section V is devoted to the study of the process of monomer-monomer catalysis reaction on
an inhomogeneous 
substrate, whose mathematical formulation is very close to that of the (inhomogeneous)
voter model
and in Section VI we summarize and present our conclusions.

\section{Voter dynamics in the presence of competing zealots}

The (homogeneous) voter model is an Ising-like model where a
spin (``individual''), associated to a lattice site ${\bm r}$, can
have two different ``opinions'' $\sigma_{\bm r}=\pm 1$ \cite{Liggett}.
The dynamics of such system is implemented by randomly choosing one
spin and changing its state to the value of one of its randomly chosen
nearest neighbors. In the (homogeneous) voter model, the global magnetization is conserved
and
the dynamics is $Z_2$ symmetric (invariance under the global
inversion $\sigma_{\bm r}\rightarrow -\sigma_{\bm r}$).
The importance of the voter model stems from the fact that it is one of a very few
stochastic
many-body systems that are solvable in any dimension. It is
useful for describing the kinetics of catalytic reactions \cite{K1,Frachebourg}, 
for studying coarsening phenomena \cite{Ben,Dornic} and also serves as a prototype
model for opinion dynamics \cite{Vazquez,IVM,Vilone}.

Concepts inspired by statistical mechanics have already been  employed to some extent in
the 
last two decades to mimic social  issues \cite{Galam}. Very recently variants of the voter
model 
and  modern tools of nonequilibrium statistical physics, such as various mean-field-like 
approaches and exact methods \cite{Sid1,IVM,Slanina}, numerical simulations
\cite{SW,Vazquez,SanMiguel,Galam2}, formulation on random networks \cite{Vilone,SanMiguel} (see
also 
references therein), were used intensively to quantitatively  study further, both
mathematically 
and numerically, collective phenomena, such as the opinion formation, inspired by
socio-cultural 
situations. In this context, the voter model and its variants play a key r\^ole, as it is
often 
used as a reference model. Despite of all these efforts, voter-like models have mainly be
studied on homogeneous and/or 
translationally-invariant spatial structures.

In contrast to most of the previous works, here we study, using exact analytical methods
and numerical simulations, 
a spatially inhomogeneous voter model. It is defined on a hypercubic lattice of size
$(2L+1)^d$, where
individuals, labeled by a vector ${\bm r}$ having components $-L \leq r_i\leq L$ (with $
i=1,\dots,
d$), may interact according to the usual voter dynamics.
In addition, we now consider that there are $n$ zealots (labeled $j=1,\dots,n$), occupying
the
sites $\{{\bm a}^{j}=(a^j_{1}, \dots, a^j_{d})\}$. These agents interact with their
neighboring
spins in a biased fashion. A zealot at site ${\bm a}^{j}$ favors one of the
opinions $\epsilon_j=\pm 1$, {\it i.e.} it flips with an additional rate $\alpha_j>0$
(additional to the
usual voter rate) toward his favorite state. 
As the zealots interact effectively with all of the spins on the lattice, 
there is a competition between them aiming at ``convincing'' as many spins as possible.
Clearly, because 
the zealots perturb the dynamics locally, the system is disordered, not translationally
invariant and 
the magnetization is not conserved.

According to the spin formulation of the model, the state of the system
is described by the collection of all spins: $S\equiv \{\sigma_{\bm r}\}$. In this
language, the
dynamics of the  model is  governed by the usual voter model transition-rate 
\cite{K1,Frachebourg,Liggett} supplemented by local terms involving the zealots'
reaction. 
The spin-flip rate, $w_{\bm r}(S)\equiv w(\sigma_{\bm r}\rightarrow -\sigma_{\bm r} )$,
therefore
reads:
\begin{eqnarray}
\label{SF}
w_{\bm r}(S)=\frac{1}{\tau}\left(1-\frac{1}{2d}\sigma_{\bm r}\sum_{{\bm r'} }
\sigma_{\bm r'} \right)
+ \sum_{j=1}^{n} \frac{\alpha_j}{2}\left( 1- \epsilon_j \sigma_{ {\bm a}^{j}}\right)\delta_{{\bm
r},{\bm a}^{j}}.
\end{eqnarray}
Here the sum on right-hand side (r.h.s.) runs over the $2d$ nearest neighbors ${\bm r'}$
of site ${\bm r}$ and $\tau\equiv 1/ d >0$ defines the time scale.
The probability distribution $P(S,t)$ satisfies the master equation:
\begin{eqnarray}
\label{master}
\frac{d}{dt} P(S,t)= \sum_{\bm r}\left[
 w_{\bm r}(S^{\bm r}) P(S^{\bm r} ,t) -
w_{\bm r}(S) P(S,t) \right],
\end{eqnarray}
where the state  $S^{\bm r}$ differs from configuration
 $S$ only by the spin-flip of $\sigma_{\bm r}$.
Using the master equation (\ref{master}), in the thermodynamic limit $L\rightarrow
\infty$,
the equation of motion of the local magnetization at site ${\bm r}$, denoted by
 $S_{\bm r}(t) \equiv \sum_{S}
\sigma_{\bm r} \,P(S,t)$, reads:
\begin{eqnarray}
\label{MS}
\frac{d}{d t} S_{\bm r}(t)=\Delta_{\bm r}S_{\bm r}(t)
+ \sum_{j=1}^{n} \alpha_j  \left(\epsilon_j-S_{{\bm a}^{j}}(t)  \right)
 \,\delta_{{\bm r}, {\bm a}^{j}}.
\end{eqnarray}

Here $\Delta_{\bm r}$ denotes the discrete Laplace operator:
$\Delta_{\bm r}S_{\bm r}(t)\equiv
 - 2d S_{\bm r}(t) +\sum_{{\bm r'} } S_{\bm r'}(t)$.
We can immediately notice from (\ref{MS}) that the stationary magnetization obeys
a discrete Poisson-like equation: $\Delta_{\bm r}S_{\bm r}(\infty)
=\sum_{j=1}^{n} \alpha_j  \left(S_{{\bm a}^{j}}(\infty) -\epsilon_j  \right)
\,\delta_{{\bm r}, {\bm a}^{j}}$. There is an obvious and striking resemblance
between this equation and the well-known equation for the electrostatic potential
generated by $n$ classical point
charges located at $\{ {\bm a}^j\}$. Therefore, one may be tempted to formally identify
$S_{\bm r}(\infty)$ with an electrostatic potential and think that the problem could be
solved easily. In fact, the problem is much harder since the quantities playing the
r\^ole of charges depend themselves on the magnetization. {\it In other
words, the problem of finding the stationary magnetization is isomorphic to the
problem of determining the electrostatic potential in a discrete system where the value
of the charges depends on the potential itself.}
Because of this fact, the calculation of $S_{\bm r}(\infty)$ cannot be inferred directly 
from the results known from electrostatics and the computations have to  be carried out in a
self-consistent manner, as described hereafter.

\section{General set-up and formal solution}

In this section, we show how to compute the magnetization of the voter model in the
presence of an arbitrary number of inhomogeneities (competing zealots) and
provide a ``formal" solution of Eq. (\ref{MS}).

For further use, we introduce the following quantity: ${\hat I}_{{\bm r}}(s)\equiv
\int_{0}^{\infty} dt\; e^{-st} \left[
e^{-2d t}I_{r_1}(2 t)\dots I_{r_d}(2 t)\right]={\hat I}_{{-\bm r}}(s)$,
where $I_n(2t)=I_{-n}(2t)=\int_0^{\pi} \frac{dq}{\pi} \, \cos{(qn)} \;e^{2t\cos{q}}$ is
the usual modified Bessel function  of first kind \cite{Abramowitz}.
The quantity ${\hat I}_{{\bm r}}(s)$ can be rewritten in terms of Watson integrals,
or ``lattice Green-functions'':
\begin{eqnarray}
\label{Wat}
{\hat I}_{{\bm r}}(s)= {\hat I}_{-{\bm r}}(s)= \int_{-\pi}^{\pi}
 \frac{d^{d}{\bm q}}{(2\pi)^d}
\frac{e^{-i\bm q\bm.\bm r}}{s+2[d-\sum_{i=1}^{d}\cos{q_i}]},
\end{eqnarray}
where ${\bm q}=(q_1,\dots,q_d)$ is a $d-$dimensional vector.
We also introduce the Fourier transform of the magnetization
\begin{eqnarray}
\label{Fourier}
{\cal S}_{\bm q}(t)=\sum_{\bm r} e^{i\bm q \bm.\bm r } \, S_{\bm r}(t).
\end{eqnarray}
Fourier transforming (\ref{MS}), we obtain the following equation:
\begin{eqnarray}
\label{FourierMS}
\frac{d}{dt}{\cal S}_{\bm q}(t) &=& -2d\, \left(
1-\frac{1}{d} \sum_{i=1}^{d} \cos{q_i}
\right)\,{\cal S}_{\bm q}(t) \nonumber\\ &+& \sum_{j=1}^{n} e^{i {\bm q \bm. \bm a}^{j}}
A^j(t),
\end{eqnarray}
where $A^j(t)\equiv \alpha_j \left(\epsilon_j - S_{{\bm a}^{j}}(t) \right).$
Laplace-transforming Eq. (\ref{FourierMS}), we obtain
 the following expression for the Laplace-Fourier transform of the magnetization:

\begin{eqnarray}
\label{LaplaceFourierMS}
{\hat {\cal S}}_{\bm q}(s)=\frac{\sum_j e^{i {\bm q \bm. \bm a}^j}\, {\hat
A}^j(s)}{s+2d\left\{ 1-\frac{1}{d}\sum_{i=1}^{d} \cos{q_i}\right\}},\end{eqnarray}
where ${\hat A}^j(s)\equiv \int_{0}^{\infty} dt\, e^{-st} A^{j}(t) $.
For technical simplicity, we have considered that the system is initially in a state with
zero magnetization: $S_{\bm r}(0)=0$.
Inverse Fourier transforming Eq. (\ref{LaplaceFourierMS}), we get the Laplace transform
${\hat S}_{\bm r} (s)$ of the magnetization:
\begin{eqnarray}
\label{LaplS}
\hat{S}_{\bm r}(s) =\sum_{\ell} \int_{-\pi}^{\pi} \frac{d^{d}{\bm q}}{(2\pi)^d}\frac{{\hat
A}^{\ell}(s) \,
e^{i({\bm a}^\ell - {\bm r}){\bm . \bm q}}}{s+2d\left\{ 1-\frac{1}{d}\sum_{i=1}^{d}
\cos{q_i}\right\}}
\end{eqnarray}
As both right and left hand-side (l.h.s.) still depend on the Laplace transform of
the magnetization (through ${\hat A}^{j}(s)$ on the l.h.s.), to obtain an explicit
expression for ${\hat S}_{\bm a^{j}}(s)$, we have to find a self-consistent solution of
Eq.(\ref{LaplS}) for all of the ${\bm a}^{j}$'s by plugging ${\bm r}={\bm a}^{j}$ into
Eq.(\ref{LaplS}).
Solving the resulting linear system, in thermodynamic limit ($L\to \infty$) we obtain:
\begin{eqnarray}
\label{Self}
\hat{S}_{{\bm a}^j}(s)
= \sum_{\ell} \int_{-\pi}^{\pi} \frac{d^{d}{\bm q}}{(2\pi)^d}
\frac{{\hat A}^{\ell}(s) \, e^{i({\bm a}^\ell - {\bm a}^j){\bm . \bm q}}}
{s+2d\left\{ 1-\frac{1}{d}\sum_{i=1}^{d} \cos{q_i}\right\}},
\end{eqnarray}
which can be rewritten $\sum_{\ell}\left({\cal M}_{j,\ell} +
\frac{\delta_{j,\ell}}{\alpha_j}\right)\, {\hat A}^{\ell}(s) = \epsilon_j/s$,
where the symmetric $n\times n$ matrix ${\cal M}$ is defined as follows:

\begin{eqnarray}
\label{Mcont}
{\cal M}_{j,\ell} (s) &=& \int_{-\pi}^{\pi} \frac{d^{d}{\bm q}}{(2\pi)^d}
\frac{e^{i({\bm a}^\ell - {\bm a}^j){\bm. \bm q}}}{s+2d\left\{ 1-\frac{1}{d}\sum_{i=1}^{d}
 \cos{q_i}\right\}}
\nonumber\\
&=& {\hat I}_{{\bm a}^j - {\bm a}^\ell}(s)= {\hat I}_{{\bm a}^\ell - {\bm a}^j}(s).
\end{eqnarray}
To obtain the two last equalities, we used the integral representation (\ref{Wat}).
We now introduce another symmetric $n\times n$ matrix, ${\cal N}$, defined by:
\begin{eqnarray}
\label{N}
{\cal N}_{j,\ell} (s,\{\alpha\})\equiv {\cal M}_{j,\ell}
(s)+\frac{\delta_{j,\ell}}{\alpha_j},
\end{eqnarray}
and from it, using Eq.(\ref{Self}), one obtains ${\hat A}^j$ and ${\hat S}_{\bm a^{j}}$:
\begin{eqnarray}
\label{AjSaj}
{\hat A}^j (s)&=& \frac{1}{s}\sum_{\ell}\epsilon_{\ell}\,[{\cal N}^{-1} (s,
\{\alpha\})]_{j,\ell} \\
 \label{AjSaj1}
{\hat S}_{{\bm a}^j} (s)&=&\frac{1}{s}\, \left(\epsilon_j
-\frac{1}{\alpha_j} \sum_{\ell} \epsilon_{\ell}\,[{\cal N}^{-1} (s,
\{\alpha\})]_{j,\ell}\,
\right)
\end{eqnarray}
At this point, we can get an explicit expression for the Laplace transform of the
magnetization by  plugging  back (\ref{AjSaj}) into (\ref{LaplS}).
In the thermodynamic limit ($L\to \infty$), we have:
\begin{eqnarray}
\label{Srcont0}
\hat{S}_{\bm r}(s) =\frac{1}{s}\sum_{j,\ell}\, \epsilon_\ell {\hat I}_{{\bm a}^j -{\bm r}}
(s)[{\cal N}^{-1} (s,\{\alpha\})]_{j,\ell},
\end{eqnarray}
and therefore, formally the magnetization is obtained by Laplace-inverting
Eq.(\ref{Srcont0}):
\begin{eqnarray}
\label{Sr}
&&S_{\bm r}(t) = \frac{1}{2\pi i}\, \nonumber\\ &\times&
\int_{c-i\infty}^{c+i\infty} \frac{ds}{s}\, e^{st} \sum_{j,\ell}\,
\epsilon_\ell {\hat I}_{{\bm a}^j -{\bm r}} (s)[{\cal N}^{-1}
(s,\{\alpha\})]_{j,\ell}.
\end{eqnarray}
This expression means that we have recast the problem of solving the inhomogeneous voter
model in the presence of arbitrary many inhomogeneities into a well-defined linear algebra
problem whose main, but nontrivial, analytic difficulty resides in the inversion of the matrix
${\cal N}$.
 The steady state of the magnetization for $L\to \infty$ can be directly
 inferred from Eq.(\ref{Srcont0}) as follows:
\begin{eqnarray}
\label{SrcontSS}
S_{\bm r}(\infty) = {\rm lim}_{s\to 0} \;
 \sum_{j,\ell}\, \epsilon_\ell {\hat I}_{{\bm a}^j -{\bm r}} (s)
 [{\cal N}^{-1} (s,\{\alpha\})]_{j,\ell}.
\end{eqnarray}
The exact expression for the long-time  magnetization is obtained  by Laplace-inverting
the $s\rightarrow 0$ expansion of Eq.(\ref{Srcont0}), after having subtracted the static
contribution $S_{\bm r}(\infty) /s$, and by paying due attention to the situations where
the integrals (\ref{Wat}) are divergent.
It is also worth mentioning that the properties of the modified Bessel functions
of the  first kind, $I_{r}(t)$ \cite{Abramowitz}, allow us to write a formal and implicit
 solution of Eq.(\ref{MS}) for $L\to \infty$, which reads :
\begin{eqnarray}
\label{formal}
&& S_{\bm r}(t) = \sum_{\bm k}  S_{\bm k}(0)
\prod_{i=1}^{d} \left[e^{-2t}I_{k_{i}-r_i}(2 t)\right]\nonumber\\ &+&
 \sum_j \alpha_j \int_{0}^{t}dt'  A^{j}(t-t')\prod_{i=1}^{d}
\left[e^{-2t'}I_{r_{i}-a^j_{i}}(2 t')\right].
\end{eqnarray}
To solve it explicitly for $S_{\bm r}(t)$, one has to Laplace transform (\ref{formal})
 and then solve the resulting linear system \cite{IVM}, which is equivalent to the
procedure described above. The expression (\ref{formal}) is advantageous if one is
interested in the global magnetization
 of the system.  In fact, as we consider an initially homogeneous and ``neutral'' system 
($S_{{\bm k}}(0)=0$), using Eqs (\ref{formal}), the global magnetization of the system can
be written:
\begin{eqnarray}
\label{globalmagn}
M(t)&\equiv& \sum_{{\bm k}}S_{{\bm k}}(t)=\sum_{j=1}^{n}\,\int_{0}^{t}d\tau \, A^{j}(\tau)
\nonumber\\
&=&\sum_{j=1}^{n}\, \alpha_j \int_{0}^{t}d\tau \, \left[
\epsilon_j -S_{{\bm a}^{j}}(\tau)
\right],
\end{eqnarray}
where we use the identity $\sum_{k=-\infty}^{\infty}I_{k}(t)=e^{t}$ \cite{Abramowitz}.

The situation considered here is particularly interesting when the zealots favor different
opinions and there is an effective competition occurring in the system. In this case we
expect nontrivial nonequilibrium space-dependent steady states. Of course, we can easily
check that in the presence of
one single zealot $(n=1)$ located at site ${\bm 0}$, with strength
$\alpha_1=\alpha$ and $\epsilon_1=1$, we recover the results reported in Reference
\cite{IVM}.
In  this case we simply have:
 ${\cal N}^{-1}= \alpha\,\left[\alpha {\hat I}_{{\bm 0}}(s) +1 \right]^{-1}$ and, together
with (\ref{LaplS}), we recover
$\hat{S}_{{\bm r}}(s)=\frac{\alpha \,{\hat I}_{{\bm r}}(s)}
{s(\alpha {\hat I}_{{\bm 0}}(s) +1)}$. In Ref. \cite{IVM}, one of us has shown that in low
dimensions the voter model with only one zealot evolves toward the unanimous state favored
by the inhomogeneity.
\section{The voter model in the presence of two competing zealots}
In this section we specifically consider the case where two competing zealots are present
($j=1,2$): One, with strength $\alpha_1=\alpha$,  located at site ${\bm a}^{1}={\bm 0}$ 
and the other located at site ${\bm a}^{2}={\bm x}$ with a strength $\alpha_2=\beta$. This
case is explicitly tractable and displays interesting features, which turns out to be
generic for the case $n>1$ as illustrated by numerical simulations.
For this case, we have
${\cal N}=\left(
\begin{array}{cc}
 {\hat I}_{\bm 0}(s) + \alpha^{-1}&  {\hat I}_{\bm x}(s) \\
 {\hat I}_{\bm x}(s) & {\hat I}_{\bm 0}(s) + \beta^{-1} \\
\end{array}
\right)$ for $L\to \infty$, and therefore, using Eq.(\ref{Srcont0}), we infer the
expression of the Laplace transform of the magnetization at site ${\bm r}$ :
\begin{widetext}
\begin{eqnarray}
\label{Srcont1}
\hat{S}_{\bm r}(s) &=&\frac{1}{s}\sum_{j,\ell}\,
{\hat I}_{{\bm a}^j -{\bm r}}(s) \,\epsilon_{\ell}
 (s)[{\cal N}^{-1} (s,\{\alpha\})]_{j,\ell} \nonumber\\ &=&
 \frac{\alpha\epsilon_1 {\hat I}_{{\bm r}}(s) + \beta \epsilon_2 {\hat I}_{{\bm r}-{\bm
x}}(s)
 +\alpha \beta \left\{ {\hat I}_{{\bm r}}(s) \left( \epsilon_1 {\hat I}_{{\bm 0}}(s) -
\epsilon_2 {\hat I}_{{\bm x}}(s) \right)   +
  {\hat I}_{{\bm r}-{\bm x}}(s) \left( \epsilon_2 {\hat I}_{{\bm 0}}(s) - \epsilon_1 {\hat
I}_{{\bm x}}(s) \right)
  \right\}
 }{s\, [1+ (\alpha +\beta ){\hat I}_{\bm 0}(s)+ \alpha\beta({\hat I}_{\bm 0}^2(s) -
 {\hat I}_{\bm x}^2(s)) ]},
\end{eqnarray}
\end{widetext}
where $\epsilon_{1,2}=\pm 1$.
Obviously, the inhomogeneous system with two zealots is interesting in the case when
$\epsilon_1=-\epsilon_2$. In fact, it is clear from  Ref. \cite{IVM} that in 1D and 2D the
condition $\epsilon_1=\epsilon_2$ implies that $S_{\bm r}(\infty)=\epsilon_1$. In this
situation, the long-time approach toward the unanimous steady state is $S_{\bm
r}(t)-S_{\bm r}(\infty)\simeq  {\cal A}t^{-1/2} $ in one dimension and  $S_{\bm
r}(t)-S_{\bm r}(\infty)\simeq
{\cal B}/\ln{t} $ in two dimensions. Thus, in low dimensions, when
$\epsilon_1=\epsilon_2$, only the long-time amplitudes ${\cal A}$ and ${\cal B}$ change
with respect to the case where $n=1$ and $\epsilon=\epsilon_1$ \cite{IVM}.

From now on, without loss of generality, we consider the more interesting situation when
there is a competition between the zealots: $\epsilon_1=-\epsilon_2=1$.
 Namely, the zealot at the origin favors the $+1$ opinion, whereas the zealot at site
${\bm x}$
  favors the opposite $-1$ state. In this case, Eq. (\ref{Srcont1}) simplifies as follows:
\begin{widetext}
\begin{eqnarray}
\label{Srcont}
\hat{S}_{\bm r}(s) =  \frac{\alpha {\hat I}_{\bm r}(s)- \beta
{\hat I}_{\bm r- \bm x}(s)+\alpha\beta \,
({\hat I}_{\bm r}(s)- {\hat I}_{\bm r-\bm x}(s))({\hat I}_{\bm 0}(s)+
 {\hat I}_{\bm x}(s))}{s\, [1+ (\alpha +\beta ){\hat I}_{\bm 0}(s)+ \alpha\beta({\hat
I}_{\bm 0}^2(s) -
 {\hat I}_{\bm x}^2(s)) ]}.
\end{eqnarray}
\end{widetext}

Different questions can be asked here: What is the range of influence of each zealot ? How
``efficient'' 
are the zealots ? How does the opinion of a randomly picked spin evolve with
the time, and what will be its final opinion ?
These questions will be answered in the next sections by explicit calculation of the
stationary 
magnetization and its long-time behavior.

\begin{figure}
\includegraphics[width=0.45\textwidth]{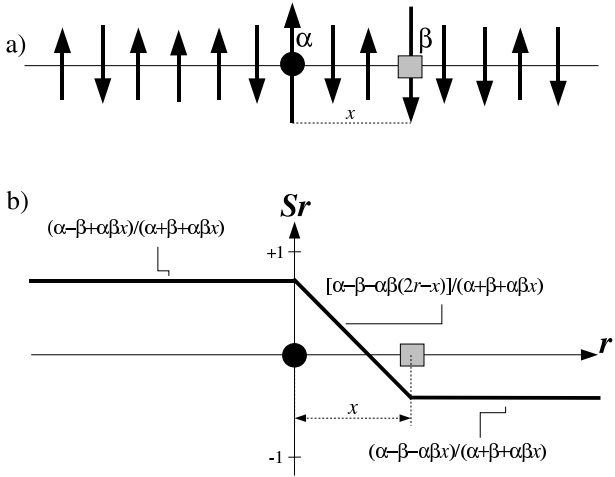}
\caption{(a) Graphical representation of a microscopic configuration of the spins on a one-dimensional
chain. The zealot favoring the $+1$ opinion with a strength $\alpha$, indicated by a dot and a 
larger up-spin, is at the origin. On the right of the origin, at a distance $x$, the
other zealot, indicated by a square and a larger down-spin, favors the $-1$ state with a
strength $\beta$. 
(b) Typical 1D stationary magnetization profile $S_{r}(\infty)$ (denoted simply $S_{r}$ in the figure)
versus $r$ in the thermodynamic limit. On the left of the origin and the right of the other zealot,
the static magnetization reaches two plateaus with heights given by Eqs. (\ref{Sr1Drfinprim}) and
(\ref{Sr1Drfinprimprim}). Between the zealots, the stationary magnetization varies linearly according to Eq.
(\ref{Sr1Drfin}).}
\label{1d_2z_sketch}
\end{figure}

\subsection{Results in 1D}
First we focus on the one-dimensional situation and
consider the case when  both competing zealots are separated by a
{\it finite} distance $x$ [See Fig.\ \ref{1d_2z_sketch}(a)]. It is worth studying the properties 
of the one-dimensional version of the inhomogeneous voter model because of its physical
implication for the catalysis (see Section V) and its close relationship with the Ising model with
Glauber dynamics, which is an important theoretical model, known to have many physical applications
\cite{Privman,IVM}. 
In fact, in the absence of zealots the one-dimensional voter model coincides with the
Glauber-Ising model with zero temperature dynamics \cite{Glauber,Muk}.

In 1D, one computes explicitly ${\hat I}_{\bm r}(s)$ in Eq. (\ref{Wat}) as follows
\cite{Abramowitz,Bateman}:
\begin{eqnarray}
\label{Ir1D}
{\hat I}_{\bm r}(s)\equiv {\hat I}_{r}(s)&=&
\frac{\left\{[\sqrt{s+4}-\sqrt{s}] /
2\right\}^{2r}}{\sqrt{s(s+4)}},
\end{eqnarray}
where $r=|{\bm r}|$.
We see that in 1D ${\hat I}_{\bm r}(s)$ diverges
 for small $s$ as $s^{-1/2}$.

Without loss of generality we consider the situation illustrated in Fig.\ \ref{1d_2z_sketch} and thus, 
from Eqs. (\ref{Srcont}), (\ref{Ir1D}),  the long-time expression for $S_r (t)$ in the case where $r\in[0,x]$ is
\begin{widetext}
\begin{eqnarray}
\label{Sr1Drfin}
  S_{r}(t)= \left(
\frac{\alpha-\beta -\alpha\beta (2r-x)}{\alpha+\beta+\alpha\beta x}
\right) 
-\frac{1}{(\alpha+\beta+\alpha\beta x)\sqrt{\pi t}}\left[
\frac{2\{\alpha-\beta-\alpha\beta(2r-x)\} +\alpha\beta x\left\{\beta(x-r)-r\alpha\right\}}{\alpha +\beta +\alpha\beta x} +\alpha r+\beta(r-x)
\right].
\end{eqnarray}
For the spins on the right of the origin, with $x<r<\infty$, we find
\begin{eqnarray}
\label{Sr1Drfinprim}
 S_{r}(t)= \left(
\frac{\alpha-\beta-\alpha\beta x}{\alpha+\beta+\alpha\beta x}
\right) -\frac{1}{(\alpha+\beta+\alpha\beta x)\sqrt{\pi t}}\left[
\frac{2(\alpha-\beta-\alpha\beta x)- \alpha^2\beta x^2}{\alpha +\beta +\alpha
\beta x}
+\alpha r+\beta(x-r)
\right],
\end{eqnarray}
whereas for the spins on the left of the origin, with $0<r<\infty$, we find:
\begin{eqnarray}
\label{Sr1Drfinprimprim}
 S_{-r}(t)= \left(
\frac{\alpha-\beta+\alpha\beta x}{\alpha+\beta+\alpha\beta x}
\right) -\frac{1}{(\alpha+\beta+\alpha\beta x)\sqrt{\pi t}}\left[
\frac{2(\alpha -\beta + \alpha\beta x) +\alpha \beta^2 x^2}{\alpha +\beta
+\alpha \beta x}
+\alpha r - \beta(r+x)
\right].
\end{eqnarray}
\end{widetext}

\begin{figure}
\includegraphics[angle=-90,width=0.45\textwidth]{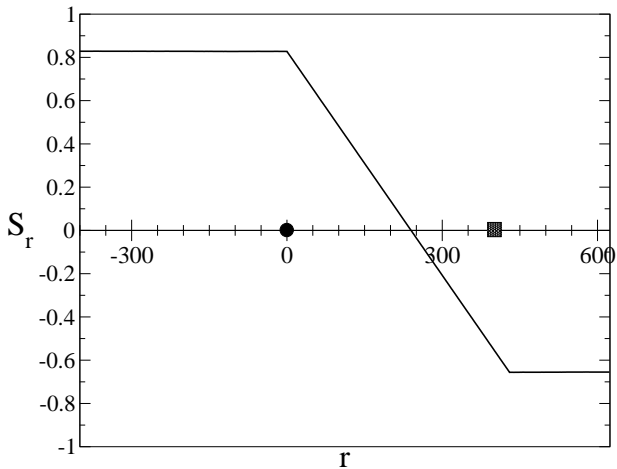}
\caption{The stationary distribution $S_r(\infty)$ on a $L=1024$ lattice with 
two competing zealots. 
The zealot favoring the positive opinion (dot) is located at the origin with $\alpha=0.02$ 
and the other one favoring the negative opinion (square) is at $r=430$ with $\beta=0.01$. The agreement
with the theoretical results for an infinite system is excellent.}
\label{Sr_1d}
\end{figure}

Finally, when both $r\to \infty$ and $t \to \infty$, ${\hat I}_{r}(s)\to
e^{-r\sqrt{s}}/(2\sqrt{s})$.
Using this expression in Eq.(\ref{Srcont}), as in Ref.\cite{IVM}, we obtain a scaling
expression for
the magnetization:
\begin{eqnarray}
\label{Sr1Dscal}
 S_{\pm r}(t)\simeq \left(
\frac{\alpha-\beta\mp \alpha\beta x}{\alpha+\beta+\alpha\beta x}
\right) \, {\rm erfc}\left(\frac{r}{2\sqrt{t}}\right),
\end{eqnarray}
where ${\rm erfc}(x)=2\int_{x}^{\infty} \frac{dy}{\sqrt{\pi}}\, e^{-y^2}$ is the usual
complementary error function.
We infer from (\ref{Sr1Drfin})  that in the finite interval separating the
two zealots, the stationary magnetization profile decays linearly with a slope 
$-2\alpha\beta/(\alpha +\beta +\alpha\beta x)$.
Outside from this interval, the final magnetization is uniform on the right and left hand side
from both inhomogeneities. In fact, (\ref{Sr1Drfinprim}) and
(\ref{Sr1Drfinprimprim}) show that the static magnetization of the spins is
$S_{\pm r}(\infty)=\frac{\alpha-\beta\mp \alpha\beta x}{\alpha+\beta+\alpha\beta x}$ (see Figs.\ 
\ref{1d_2z_sketch}(b), \ref{Sr_1d}).
These plateaus differ significantly from the values $\mp 1$ only when when the product
$\alpha\beta$ is comparable to $x^{-1}$.
Therefore in 1D, the final stationary solution, which is summarized on Fig.\ \ref{1d_2z_sketch}(b), 
is {\it polarized} and can be understood as being the solution of a discrete one-dimensional electrostatic
Poisson equation with peculiar boundary conditions.
In fact, it is well known that in $1D$ the electrostatic potential varies linearly with
the distance to the charges. Here, the nontrivial part of the analysis is to compute in a self-consistent manner
 the heights of the plateaus. 
All these profiles are approached algebraically in time, i.e.
$S_r(t)-S_r(\infty)\simeq A t^{-1/2}$ (as in the case with only one zealot \cite{IVM}),
with amplitude depending nontrivially of all parameters of the system $A={\widetilde
A}(\alpha,\beta,x) r$. Obviously, because there is a distance $x$ separating the zealot at the origin
from the other, the expression for $S_r(t)$ is not symmetric with respect to
the site $0$. We can notice that the expressions (\ref{Sr1Drfin}),(\ref{Sr1Drfinprim}) and
(\ref{Sr1Drfinprimprim}) simplify when the strength of the zealot is infinite ($\alpha=\beta=\infty$). In this case, the zealots have a final magnetization $S_0(\infty)=-S_x(\infty)=1$.  

Result (\ref{Sr1Dscal}) tells us that for spins {\it infinitely} far away from the
zealots, the magnetization evolves as a smooth scaling function of the variable 
$u\equiv \frac{r}{2\sqrt{t}}$. This scaling function differs from zero (the initial
condition) after a long time ({\it i.e.} $t\sim r^{2}\to \infty$), when the variable $u$ has a finite
value. It  follows from Eqs.(\ref{Sr1Drfinprim}),(\ref{Sr1Drfinprimprim}) and (\ref{Sr1Dscal}) that in 1D the effect of
the zealots is felt and propagates as $t^{1/2}\to \infty$. For large time and distance,
when $1 \ll t\ll r^2$, we see from Eq.(\ref{Sr1Dscal}) that $S_r (t)$ is still close to its
initial value. When $t\sim r^2$, all the agents
approach as $t^{-1/2}$ the active and fluctuating stationary magnetization
(\ref{Sr1Drfinprim}). From Eqs. (\ref{Sr1Drfin}) and (\ref{globalmagn}),
we can infer the long-time behavior of the global magnetization in the system. As 
$\alpha (1-S_{0}(t))-\beta (1+S_{x}(t))\simeq \frac{2(\alpha-\beta)}
{\alpha+\beta+\alpha\beta x}\frac{1}{\sqrt{\pi t}}$ when $\alpha\neq \beta$, the 
average number of voters following the 
strongest zealot evolves (at long-time) as the square-root of time: $M(t)\simeq  
\frac{4(\alpha-\beta)}{\alpha+\beta+\alpha\beta x}\sqrt{\frac{t}{\pi}}$. 
This result implies that the time $T$ necessary for the strongest
zealot to dominate (on average) the whole 1D system scales as  $T \sim  L^{2}$, where $L\to \infty$.
When 
$\alpha=\beta$, the system is exactly symmetric with respect to $x/2$, and in average there 
are as many $+1$ spins than $-1$ ones in the whole system.

On Fig.\ \ref{Sr_1d} we show the stationary magnetization $S_r(\infty)$ on a finite
lattice with $L=1024$ for two competing zealots obtained from Monte Carlo simulations.
For simulating the model we use random sequential dynamics by picking randomly
an ``active'' site ( either one of the zealots or a site that has at least one
nearest neighbor in a different state) and flipping it with a 
rate given by Eq. (\ref{SF}). The time after an attempt for a flip is updated with
the amount $1/N_a$, where $N_a$ is the number of active sites before the current update.
To account for the fact that the simulations are on a finite lattice, 
where the spin at the left (right) boundary site has 
only one nearest neighbor on the right (left), the spin-flip rate at the boundaries is
modified such that it depends only on the state of a single  neighbor.
The first $2\times10^8$ Monte Carlo steps (MCS) are discarded and typically we 
sample the configurations on the lattice every $5000$ MCS for the next $5\times10^9$ MCS. 
The stationary distribution for $S_r(\infty)$ obtained from the simulations is in an excellent 
agreement with the theoretical values obtained for
a infinite lattice and sketched on Fig.\ \ref{1d_2z_sketch}(b).

\begin{figure}
\includegraphics[angle=-90,width=0.45\textwidth]{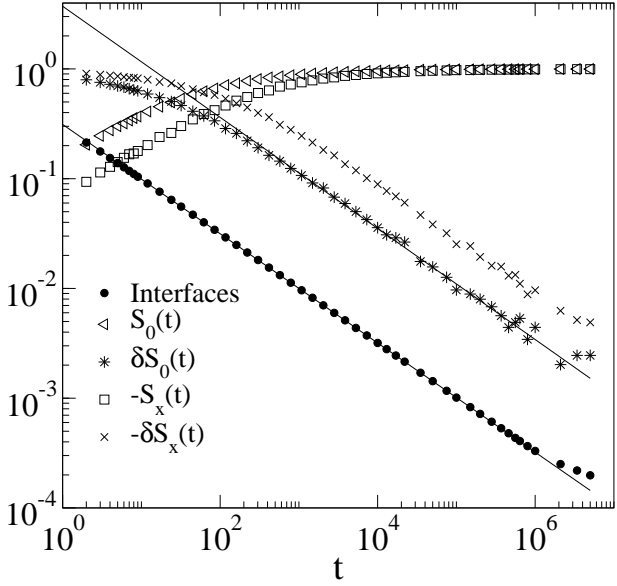}
\caption{Coarsening on the one-dimensional model with two competing zealots. The figure
shows the average number of interfaces vs. time, the average magnetization of the 
two zealots (see the text) $S_0(t)$ and $S_x(t)$, and also $\delta S_0(t) \equiv S_0(\infty) - S_0(t)$ and
$\delta S_x(t) \equiv S_x(\infty) - S_x(t)$. The simulation is on $L=8192$ lattice for
$\alpha=0.5$, $\beta=0.2$ and $x=3000$ and the continuous lines shown have a slope $-0.5$, 
as predicted by 
Eq. (\ref{Sr1Drfin}). For this choice of the parameters, the average number of interfaces 
decays algebraically toward a small but finite value (here, $\approx 2.0\times 10^{-4}$).}
\label{1d_2z}
\end{figure}

Fig.\ \ref{1d_2z} shows the result from Monte Carlo simulations on a relatively small $(L=8192)$ lattice
for various average quantities. 
The long time behavior of the local magnetization 
$\delta S_0(t)\equiv S_0(\infty) - S_0(t)$ and 
$\delta S_x(t)\equiv S_x(\infty) - S_x(t)$ clearly show the $t^{-1/2}$ long time behavior, 
in agreement with Eq.(\ref{Sr1Drfin}).
In Fig. \ \ref{1d_2z} we also report numerical results for the average number of interfaces ({\it i.e.} 
two neighboring sites with different spins) vs. time. This quantity gives us a good qualitative 
and quantitative picture of the coarsening of the system.
Fig. \ \ref{1d_2z}  shows that the average value of the interfaces, which equals to 
the number of the clusters of $+1$ and $-1$ spins, evolves as $t^{-1/2}$ before 
saturating at a small non-zero value. One can notice that for a long time the system evolves 
and {\it coarsens} as in the homogeneous  voter model \cite{Frachebourg}, but due to the presence of the 
two competing zealots, subtleties appear at long times.
In fact, one has to distinguish between the three possible situations for the
coarsening: (i) when we have $n<2$ (i.e. none or only one zealot on the lattice), there is
the usual coarsening (an infinite domain spans the entire system) \cite{Frachebourg}; (ii) when  $2\leq n$ 
and the density ($n/L$) of the competing inhomogeneities is zero for $L\rightarrow \infty$, there is
still coarsening in the sense that the size of the different domains formed increases with the
size of the lattice but never spans the entire lattice; (iii) when the density of the competing zealots has a non-zero value in the thermodynamic limit, there is no longer coarsening as the formation 
of large domains is prevented by the interaction with the numerous (competing) inhomogeneities.

After having discussed in detail the case $n=2$, we would like to point out that in one
spatial dimension it is possible to compute the stationary magnetization for an arbitrary
number $n$ of zealots in a more direct and intuitive fashion than relying on
Eq.(\ref{SrcontSS}). In fact, let us consider that the zealots, labeled by $j=1, \dots,
n$ are at sites $-\infty<a_1<a_2<\dots<a_n<\infty$. By plugging  the ansatz that the
stationary magnetization between the sites $a^{j}$  and $a^{j+1}$  reads
$S_r(\infty)=S_{a^{j}}(\infty)+\gamma_j (r-a^j)$ into
$\Delta_{ r}S_{ r}(\infty)=-\sum_{j=1}^{n} \alpha_j  \left(\epsilon_j-S_{a^{j}}(\infty)
\right)
 \,\delta_{r, a^{j}}$, where we have introduced $\gamma_j\equiv \frac{S_{a^{j+1}}(\infty)
-S_{a^{j}}(\infty)}{x_j}$ and $x_j\equiv a^{j+1}-a^j $, we obtain:
\begin{eqnarray}
&&\gamma_1 \delta_{r,a_1} + (\gamma_2 -\gamma_1)\delta_{r,a_2} + \dots + (\gamma_{n-1}
-\gamma_{n-2})\delta_{r,a_{n-1}} \nonumber\\ &-&\gamma_n \delta_{r,a_n} =\sum_{j=1}^{n}
\alpha_j  \left(S_{a^{j}}(\infty) -\epsilon_j \right)\,\delta_{r, a^{j}}.
\end{eqnarray}
Solving these equations, we obtain the stationary magnetization at each sites $a^1\leq a^j
\leq a^n$:
\begin{eqnarray}
\label{Saj}
S_{a^{1}}(\infty)&=&\epsilon_1 + \frac{\gamma_1}{\alpha_1} \nonumber\\
S_{a^{2}}(\infty)&=&\epsilon_{2} +\frac{\gamma_2 -\gamma_1}{\alpha_{2}}\nonumber\\
\vdots \quad &\vdots& \quad\vdots \nonumber\\
S_{a^{n-1}}(\infty)&=&\epsilon_{n-1} +\frac{\gamma_{n-1}
-\gamma_{n-2}}{\alpha_{n-1}}\nonumber\\
S_{a^{n}}(\infty)&=&\epsilon_{n} -\frac{\gamma_{n-1}}{\alpha_{n}}
.
\end{eqnarray}
Of course, in each of these equations for $S_{a^j}(\infty)$, the right-hand-side depends
on $S_{a^j}(\infty)$
and $S_{a^{j+1}}(\infty)$ through $\gamma_j$. The equations (\ref{Saj}) are therefore a set
of coupled linear equations that can be rewritten as $P {\bm S}= {\bm v}$, where $P$ is a
$n \times n$ band matrix, which only non-vanishing entries are
\begin{eqnarray}
\label{P}
P_{j,j}&=&-(x_{j-1}+ x_{j}+\alpha_j x_{j-1}x_j), \quad 1<j<n \nonumber\\
P_{j,j-1}&=& x_j, \quad 1<j<n  \nonumber\\
P_{j,j+1}&=& x_{j-1}, \quad 1<j<n  \nonumber\\
P_{1,1}&=&-(1+\alpha_1 x_{1}) \nonumber\\
P_{n,n}&=&-(1+\alpha_n x_{n-1}) \nonumber\\
P_{1,2} &=& P_{n,n-1}=1,
\end{eqnarray}
and ${\bm S}$ and ${\bm v}$ are column vectors which components are respectively
\begin{eqnarray}
\label{Sandv}
S_j &=& S_{a^j}(\infty) , \quad 1\leq j \leq n \nonumber\\
v_{1} &=&-\epsilon_1 \alpha_1 x_{1}\nonumber\\
 v_{j} &=& -\epsilon_j \alpha_j x_{j-1}x_j, \quad 1<j< n \nonumber\\
 v_{n} &=& -\epsilon_n \alpha_n x_{n}.
\end{eqnarray}

\begin{figure}
\includegraphics[angle=-90,width=0.45\textwidth]{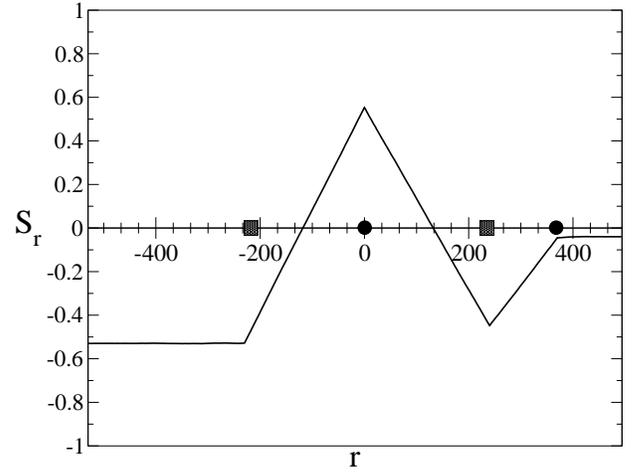}
\caption{An example for numerical simulation of the case with $4$ zealots on a $L=1024$ lattice (see the text).
The bias of the zealots from left to right is $0.01$ (negative), $0.02$ (positive), $0.013$ (negative)
and $0.003$ (positive).}
\label{1d_4z}
\end{figure}

Therefore, the solution of (\ref{Saj}) is obtained by inverting the band matrix $P$  and
reads:
\begin{eqnarray}
\label{SajSol}
S_{a^j}(\infty)=\sum_{k=1}^{n}[P^{-1}]_{j,k} \, v_k.
\end{eqnarray}
Having solved (at least formally) the set of equations (\ref{Saj}) giving the stationary
magnetization at each site $a^j$, the general one-dimensional stationary magnetization in
the presence of $n$ zealots simply reads:
\begin{itemize}
\item If $r<a^1$:
\begin{eqnarray}
\label{magnstat_arb1}
S_{r}(\infty)= S_{a^{1}}(\infty).
\end{eqnarray}
\item If $a^j\leq r\leq a^{j+1}$ ($1 \leq j<n$):
\begin{eqnarray}
\label{magnstat_arb2}
S_{r}(\infty)= S_{a^{j}}(\infty)+\frac{S_{
a^{j+1}}(\infty)-S_{a^{j}}(\infty)}{a^{j+1}-a^j}(r-a^j).
\end{eqnarray}
\item If $r>a^n$:
\begin{eqnarray}
\label{magnstat_arb3}
S_{r}(\infty)= S_{a^{n}}(\infty).
\end{eqnarray}
\end{itemize}

As an example, let us consider the case where there are four zealots on the chain, as illustrated in Fig.\ \ref{1d_4z}. This figure shows that the one-dimensional stationary magnetization profile is a
piecewise function, as predicted by Eqs.(\ref{magnstat_arb1})-(\ref{magnstat_arb3}).
When $n=4$, as in Fig.\ \ref{1d_4z}, Eqs. (\ref{Saj}) explicitly read :
\begin{eqnarray}
\label{4Z}
 S_{a^2}(\infty)&-&(1+\alpha_1x_1)S_{a^1}(\infty)=-\epsilon_1\alpha_1 x_1 \nonumber\\
 x_1S_{a^3}(\infty)&-&(x_1 +x_2+ \alpha_2 x_1 x_2) S_{a^2}(\infty) + x_2
S_{a^1}(\infty)\nonumber\\ &=&-\epsilon_2\alpha_2 x_1 x_2 \nonumber\\
x_2 S_{a^4}(\infty)&-&(x_2 +x_3+ \alpha_3 x_2 x_3) S_{a^3}(\infty) + x_3
S_{a^2}(\infty)\nonumber\\ &=&-\epsilon_3\alpha_3 x_2 x_3\nonumber\\
 S_{a^3}(\infty)&-&(1+\alpha_4 x_3)S_{a^4}(\infty)=-\epsilon_4\alpha_4 x_3
\end{eqnarray}

The set of Eqs.(\ref{4Z}) can be solved explicitly and gives rise to very cumbersome
expressions.
Plugging into the latter the values corresponding to the system simulated in Fig.\ \ref{1d_4z},
{\it i.e.} $\alpha_1=0.01, \epsilon_1=-1$,  $\alpha_2=0.02, \epsilon_2=+1$,  $\alpha_3=0.013,
\epsilon_3=-1$ and  $\alpha_4=0.003, \epsilon_4=+1$, and $x_1=230, x_2=240, x_3=130$, we
obtain:
$S_{a^1}(\infty)= -0.529, S_{a^2}(\infty)= +0.556, S_{a^3}(\infty)= -0.441,
S_{a^4}(\infty)= -0.0367$.
These values can be compared to the results of the simulations, reported in Fig.\ \ref{1d_4z} , where
we obtained
$S_{a^1}(\infty)= -0.53 \pm 0.01, S_{a^2}(\infty)= +0.55\pm 0.01, S_{a^3}(\infty)=-0.45\pm
0.01, S_{a^4}(\infty)=-0.04\pm 0.005$.
These comparisons show that there is an excellent agreement between the theoretical values
predicted by the solution (\ref{SajSol}) of the system (\ref{4Z}) and the numerical results. This agreement
is somewhat surprising as the simulations reported in Fig.\ \ref{1d_4z} have been carried on a
relatively small system ($L=1024$), whereas all the theoretical results
(\ref{Saj})-(\ref{4Z}) have been derived in the thermodynamic limit. This fact indicates
that our analytic results may be quantitatively accurate even for large, but non-infinite, systems. In the limit where the strength of the zealots is $\alpha_1=\dots=\alpha_n=\infty$, all the expressions simplify and it folows from (\ref{Saj}) that $S_{a^j}(\infty)=\epsilon_j$, 
while, for $a^j \leq r \leq a^{j+1}$,
$S_r(\infty)= \epsilon_j+\left(\frac{\epsilon_{j+1}-\epsilon_{j}}{a^{j+1}-a^j}\right)\,(r-a_j)$.
When $\alpha_1=\dots=\alpha_n=\infty$, this 1D system can be related to the one-dimensional spin model with Glauber dynamics (at zero-temperature) in the presence of quenched random fields of infinite strength \cite{Muk}: in the voter language, the situation considered by the authors of Ref.\cite{Muk} would correspond to the case where at each site $j$ a ``voter'' would have a probability $p$ to be a zealot favoring the opinion $\epsilon_j =\pm 1$ with strength $\alpha_j=\infty$ and would have a probability $1-2p$ to be an ordinary agent. The (slight) difference between such a model and the one studied in Ref.\cite{Muk} is the fact that  each zealot (even when he is endowed with an infinite strength) can be ``forced'' to flip by his two neighbors, while in Ref.\cite{Muk} the (random)
 magnetic fields pin the spins along their direction. However, as $\alpha_j=\infty$, each zealot 
 $j$ rapidly flips back to his preferable opinion $\epsilon_j$ and thus both models are very close and display the same stationary magnetization. 

We also would like to emphasize that the results (\ref{Saj}), (\ref{SajSol}) provide the
exact magnetization of the completely disordered one-dimensional voter-model, where each site is
endowed with a specific spin-flip rate. In this case, one would have  $n=L\to \infty$ zealots in
the system with $x_j=a^{j+1}-a^j=1$, and the structure of the matrix $P$ is rather simple [see Eq.(\ref{P})].

\subsection{Results in 2D}

\begin{figure}[!t]
\includegraphics[width=0.45\textwidth]{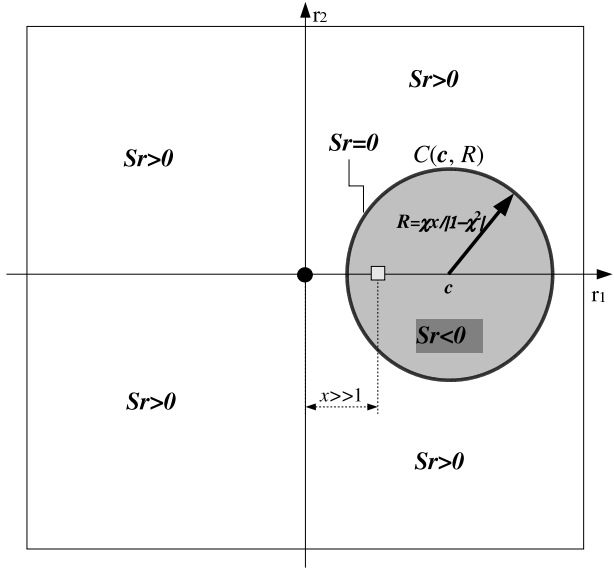}
\caption{Sketch of the typical 2D spatial dependence of the stationary magnetization when
$L\to \infty$. At the origin, indicated by a dot, is the zealot favoring the state $+1$ with a strength
$\alpha=1$. At a distance $x\gg 1$, indicated by a square, is the zealot favoring the state $-1$ with
a strength $\beta\simeq 0.9$. According to Eq. (\ref{Sr2dstat}), the agents within the disk of center ${\bm c}\simeq 2{\bm x}$ and of radius $R\simeq 1.4x$ have a negative final magnetization (denoted simply
$S_{\bm r}$ in the figure). Outside the disk, the final magnetization of the
agents is positive (see the text), while on the circle  the agents are in a ``neutral''
final state. The static magnetization $S_{\bm r}(\infty)$ exhibits both radial and polar dependence.}
\end{figure}

In two dimensions, the integral of Eq.(\ref{Wat}) is also divergent
in the long-time regime $s\rightarrow 0$ and therefore its main contribution arises from
$ q^2\equiv q_1^2 +q_2^2 \rightarrow 0$.
In this sense, we first expand Eq.(\ref{Wat}) for small $s$
 in the case when ${\bm r}={\bm 0}$:
\begin{eqnarray}
\label{I02d}
{\hat I}_{ \bm 0}(s) \xrightarrow[s\rightarrow 0]{}
-\frac{1}{4\pi}\ln{s},
\end{eqnarray}

More generally, for $r\gg 1$, we have (see Ref.\cite{IVM}) ${\hat I}_{ \bm r}(s)
\xrightarrow[s\rightarrow 0]{}
\frac{1}{2\pi}K_{0}\left(r\sqrt{s}\right)$, where  $K_{0}(x)$ is the usual modified Bessel
 function of the third kind \cite{Abramowitz}.
Using the small argument expansion of such a
 Bessel function
we find that the long-time behavior for $t\gg r^{2}\gg 1$ is given by
\begin{eqnarray}
\label{Ir2d}
{\hat I}_{ \bm r}(s)
\xrightarrow[r\sqrt{s}\rightarrow 0]{}
-\frac{1}{4\pi} \, \left[\ln{(r^2 s)} + 2\{\gamma - \ln{2}\}\right],
\end{eqnarray}
where $\gamma=0.5772156649 \dots$ denotes the usual Euler-Mascheroni's constant.
From the expression (\ref{Srcont}), when $x$ is sufficiently large to use
Eq.(\ref{Ir2d}), we obtain the stationary magnetization of the zealots:
$S_{{\bm 0}}(\infty)\simeq\frac{\alpha -\beta +\frac{\alpha
\beta}{\pi} \ln{x}}{\alpha +\beta +\frac{\alpha
\beta}{\pi} \ln{x}}$ and $S_{{\bm x}}(\infty)\simeq\frac{\alpha -\beta -\frac{\alpha
\beta}{\pi} \ln{x}}{\alpha +\beta +\frac{\alpha
\beta}{\pi} \ln{x}}$. Interestingly these expressions resemble to the ones obtained
in 1D [see Eqs (\ref{Sr1Drfinprim}), (\ref{Sr1Drfinprimprim})]. The only change is in the
dependence on
separating distance: With respect to the 1D case, one has $x \to \frac{1}{\pi}\ln{x}$.
When $r\gg 1$ and $|{\bm r-\bm x}|  \gg 1$, from (\ref{Srcont}), using Eqs.(\ref{I02d})
and (\ref{Ir2d}), 
the stationary magnetization reads (see Fig. 5):
\begin{eqnarray}
\label{Sr2dstat}
S_{\bm r}(\infty)\xrightarrow[ r\gg 1, |{\bm r-\bm x}| \gg 1]{}
\frac{\alpha - \beta -\frac{\alpha\beta}{\pi} \, \ln{\frac{r}{|{\bm r- \bm x}|}} }
{\alpha + \beta +\frac{\alpha \beta}{\pi}\,
[\ln{x}+\pi(\gamma-\ln{2})] }
\end{eqnarray}

\begin{figure*}
\includegraphics[width=6in]{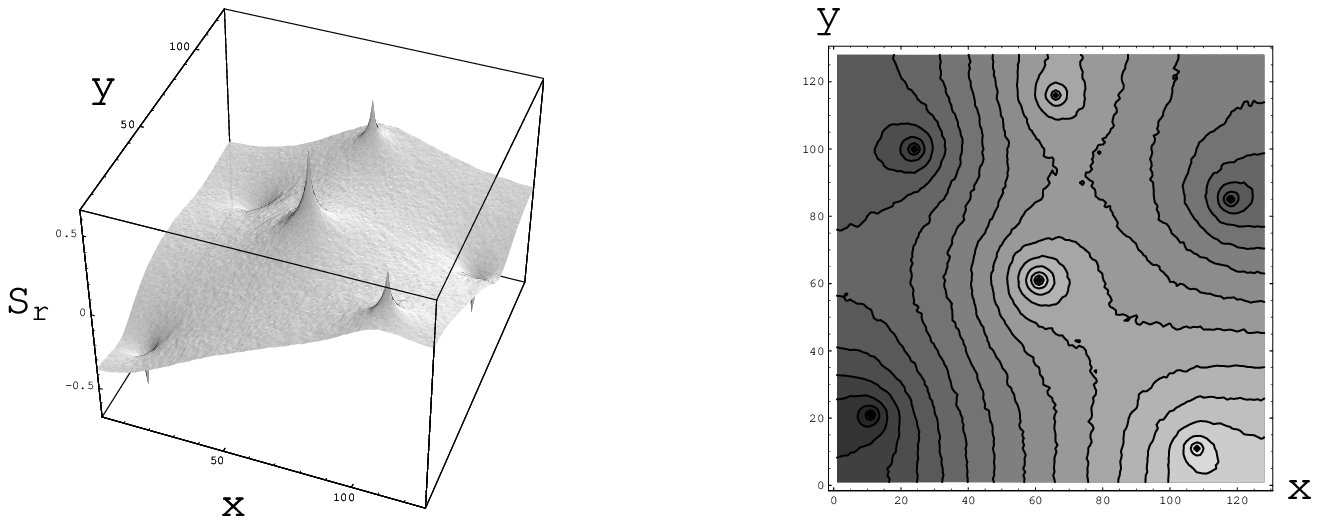}
\caption{The stationary site magnetization $S_{\bm r}(\infty)$ on a $(128\times128)$ lattice in the
presence of six zealots. Three of the zealots favor the $+1$ state and other three the $-1$ opinion.
The picture on the left shows a 3D plot of $S_{\bm r}(\infty)$ (along the vertical
axis) and the picture on the right is the corresponding contour plot. The strengths
of the positive zealots are $2.0, 1.2, 0.8$ and the strengths of the negative ones are
$1.6, 1.4, 1.0$.}
\label{2d_six_z}
\end{figure*}

Far away from both zealots, and in the case of sufficiently separated zealots, {\it i.e.}
$r \gg x  \gg 1$, this expression simplifies:
\begin{eqnarray}
\label{Sr2dst}
S_{\bm r}(\infty)\xrightarrow[r\gg x\gg 1]{} \frac{\alpha - \beta -\frac{\alpha\beta}{\pi}
\frac{x}{r}\cos{\theta}}{\alpha + \beta +\frac{\alpha \beta}{\pi}\,
[\ln{x}+\pi(\gamma-\ln{2})] },
\end{eqnarray}
where $\cos{\theta}\equiv \frac{{\bm r \bm. \bm x}}{r x}$. Here we used
the fact that $\ln(r/|{\bm r- \bm x}|)=\frac{x \cos{\theta}}{r} + {\cal O}((x/r)^2)$,
when $r\gg x\gg 1$. These results show that, because of the competition between the two
zealots, the 
stationary magnetization is a fluctuating steady state exhibiting nontrivial radial and
polar dependence. Also, when $\alpha=\beta=\infty$, Eq.(\ref{Sr2dstat}) reduces to
$S_{\bm r}(\infty)\xrightarrow[ r\gg 1, |{\bm r-\bm x}| \gg 1]{}
\frac{\ln{\frac{|{\bm r- \bm x}|}{r}} }
{\ln{x}+\pi(\gamma-\ln{2})}$ and $S_{\bm 0}(\infty)=-S_{\bm x}(\infty)=1$.

Regarding the dynamical behavior, in the regime where $t\gg {\rm max}(|{\bm r}-{\bm
x}|^{2}, r^2)$, the 
long-time behavior of the magnetization is the following:
\begin{widetext}
\begin{eqnarray}
\label{Sr2ddyn}
S_{\bm r}(t) - S_{\bm r}(\infty) \simeq
-\frac{1}{\ln{t}}\, \left[
\frac{\ln{\frac{r^{2\alpha}}{|{\bm r-\bm x}|^{2\beta}}} -
\frac{\alpha \beta}{\pi}\, \ln{\frac{r}{|{\bm r-\bm x}|} \{\ln(x/2) + \gamma\}}
+2(\alpha -\beta)(\gamma -\ln{2})
}{
\alpha + \beta +\frac{\alpha \beta}{\pi}\,
\{\ln{x}+\pi(\gamma-\ln{2})\}
 }
\right].
\end{eqnarray}
\end{widetext}

In the situation where $r\gg x\gg 1$, the above expression simplifies and the approach
toward the steady-state (\ref{Sr2dst}) is $S_{\bm r}(t) - S_{\bm r}(\infty) \simeq
-\frac{1}{\ln{t}}\, \left[
\frac{2[(\alpha-\beta)\ln{r}+\frac{x}{r}\beta\cos{\theta}]
-{\frac{\alpha\beta}{\pi}}
\{\frac{x}{r} \cos{\theta}\ln{x}\}
}{
\alpha + \beta +\frac{\alpha \beta}{\pi}\,
\{\ln{x}+\pi(\gamma-\ln{2})\}
 }
\right]$. For $t\gg r^2$, these results tell us that the 2D system evolves logarithmically
slowly toward a space-dependent fluctuating steady state.
As in the presence of only one zealot, we can see that in 2D the magnetization does not
exhibit a scaling expression between $r$ and $t$ when $r^2 \sim t \gg 1$ \cite{IVM}.
This is due to the logarithmic terms, specific to the two-dimensional situation, appearing
in (\ref{I02d}) and (\ref{Ir2d}). Natural questions arise regarding the spatial
distribution of ``opinions'': {\it  What is the spatial voting distribution in the 
steady-state? Which region is characterized by a
majority of positive/negative opinion ? How does the strength of $\alpha$ and $\beta$
affect the final spatial opinion distribution ?}

To answer these questions, we use Eq.(\ref{Sr2dstat}) and notice that in the limit $r\gg
1$ and $|{\bm r- \bm x}|\gg 1$,  the spatial region  where $S_{\bm r}(\infty)=0$ obeys the
equation:
\begin{eqnarray}
\label{reg0magn}
\frac{r}{\mid {\bm r-\bm x} \mid}=\chi^{-1} \; \quad \text{with} \quad
\chi \equiv {\rm exp}\left(\frac{\pi[\beta -\alpha]}{\alpha \beta}\right)
\end{eqnarray}
When $\alpha \neq \beta$, i.e. for $\chi\neq 1$,
 such an equation can be recast into the following form:
$r^2+\frac{2rx}{\chi^2 -1}\, \cos{\theta} - \frac{x^2}{\chi^2 -1}=0$, {\it i.e.} the polar
equation of a circle ${\cal C}({\bm c}, R)$ centered at
${\bm c}=\frac{1}{1-\chi^2}{\bm x}$ and of radius $R=\frac{\chi x}
{\mid 1-\chi^2\mid} = x/2\sinh{\left(|\alpha^{-1} - \beta^{-1}|\right)}$.
 This result, together with (\ref{Sr2dstat}) and (\ref{reg0magn}), implies
that in 2D, for $\alpha \neq \beta$, the agents located on
the circle ${\cal C}({\bm c}, R)$ are ``neutral'' they have
zero final magnetization as illustrated in Fig. 5.
From (\ref{Sr2dstat}) and (\ref{reg0magn}) we can also conclude that:

$\bullet$ If $\chi>1$, {\it i.e.} $\beta>\alpha$, the agents that are within
(outside) the disk ${\rm Int}\,{\cal C}({\bm c}, R)$ have a positive
(negative) magnetization.

$\bullet$ If $\chi<1$, {\it i.e.} $\beta<\alpha$, the agents that are
within (outside) the disk ${\rm Int}\,{\cal C}({\bm c}, R)$ have a negative
(positive) magnetization. This case is sketched in Figure 2.

These results show that the majority of the voters, except the ones enclosed in
the disk, tend to follow the opinion favored by
the strongest zealot. The details of the neutral region
${\cal C}({\bm c}, R)$ depend nontrivially on all the parameters $\alpha, \beta$ and
${\bm x}$ and, interestingly, the radius grows with the difference of the strength 
of the zealots as $R\propto 1/\sinh{u} $, where
$u\equiv\beta^{-1}-\alpha^{-1}$. Also, $R$  increases linearly with the separating distance
$x$.

$\bullet$ The case $\alpha=\beta$ (including $\alpha=\beta=\infty$), {\it i.e.} $\chi=1$, is special.
In this situation, it follows from (\ref{reg0magn}) that the region with zero-final
magnetization is no longer a closed curve but an infinite line given by the equation
$r=\frac{x}{2\cos{\theta}}$ which separates the two-dimensional space into two
semi-infinite half-planes.

For the number of zealots $n>2$ the analytical calculations become very tedious and
we illustrate the results of a Monte Carlo simulation of the case with six zealots on 
Fig.\ \ref{2d_six_z}. The simulation is carried on a $(128\times128)$ lattice and due
to the $1/\ln(t)$ approach to the steady state enormous sampling times are required.
Again when simulating the system one has to be careful with the sites on the boundaries:
if the site lies on the edges then it has only three nearest neighbors; and if it is at
the corners, then it has only two nearest neighbors. The stochastic rules have to be slightly 
modified to account for the boundary sites.
The geometry of the zealots can be seen from contour plot on 
Fig.\ \ref{2d_six_z} where three of the zealots are positively biased and three 
are biased negatively. The left picture on Fig.\ \ref{2d_six_z} shows the average
magnetization $S_{\bm r}(\infty)$ on the different sites of the lattice. For these particular
values of the bias of the zealots and their position on the lattice, in the stationary state,
 we observe one large region of
positive on average opinion (a curved central ``stripe'' in Fig.\ \ref{2d_six_z}) and two smaller disconnected regions of a negative opinion (near the left boundary and top right edge of  Fig.\ \ref{2d_six_z}).

 Regarding the coarsening of the 2D system, we again distinguish three situations: (i)
when $n<2$, there is usual coarsening and an infinite domain eventually spans the entire system; 
(ii) when there is a finite number of competing zealots large domains still
develop but their size is limited by the zealots; (iii)
 when the density of the competing zealots is finite
in the thermodynamic limit, there is no longer coarsening as the formation 
of large domains is prevented by the interaction with the numerous inhomogeneities.

To conclude this section, as in 1D, we notice that
$\alpha(1-S_{\bm 0}(\infty))=\beta(1+S_{\bm x}(\infty))$ which implies, with 
Eq.(\ref{globalmagn}), that the global magnetization evolves, following the strongest zealot ($\alpha\neq \beta$), as $M(t)\sim t/\ln{t}$.  As a consequence, the time $T$ necessary for the strongest
zealot to dominate (on average) the whole 2D system is  $T \sim L^{2} \ln{L} $ (where $L\to \infty$).
In the symmetric case ($\alpha = \beta$), as explained above, the 2D space is exactly separated in two
 semi-infinite half-planes with opposite total magnetization. 

\subsection{Results in 3D}

Above two dimensions, the integrals in Eq. (\ref{Wat}) are well defined for all values of
$s$ and in particular when $s\rightarrow 0$. Therefore, in contrast to what happens in
$1D$ and $2D$, to determine the long-time behavior of the
 magnetization we cannot simply focus on the $q\rightarrow 0 $ expansion of (\ref{Wat}).
 This also means that in dimensions $d\geq 3$ in the presence of $n$ zealots
  the static magnetization readily follows from from Eq.(\ref{SrcontSS}):
\begin{eqnarray}
\label{SrSS3D}
S_{\bm r}(\infty) =  \sum_{j=1}^{n} \sum_{\ell=1}^{n}\, \epsilon_\ell {\hat I}_{{\bm a}^j
-{\bm r}} (0)
 [{\cal N}^{-1} (0,\{\alpha\})]_{j,\ell}.
\end{eqnarray}
The three-dimensional lattice Green function ${\hat I}_{{\bm r}} (0)$ has been computed
very recently by Glasser and Boersma \cite{Glasser}. Using the triplet
 $(a_{{\bm r}}, b_{{\bm r}}, c_{{\bm r}})$ of rational numbers
depending on ${\bm r}$, given in Table 2 of Reference
\cite{Glasser}, and the quantity
$g_0\equiv\left(\frac{\sqrt{3}-1}{96\pi^{3}}\right)\Gamma^{2}
\left(\frac{1}{24}\right)\Gamma^{2}\left(\frac{11}{24}\right)=0.505462...$ ($\Gamma(z)$ is
Euler's Gamma function), it has been established that:
\begin{eqnarray}
\label{Ir3D}
{\hat I}_{\bm r}(0)= \frac{1}{2}
\left[a_{\bm r} g_0 + c_{\bm r}  +\frac{b_{\bm r} }{\pi^{2}g_0}
\right].
\end{eqnarray}
With (\ref{Srcont}) and  (\ref{Ir3D}) the exact expression of the three-dimensional
magnetization
in the presence of two zealots is explicitly given by:
\begin{widetext}
\begin{eqnarray}
\label{Sr3DSTAT}
S_{\bm r}(\infty) = \frac{\alpha\epsilon_1 {\hat I}_{{\bm r}}(0) + \beta \epsilon_2 {\hat
I}_{{\bm r}-{\bm x}}(0)
 +\alpha \beta \left\{ {\hat I}_{{\bm r}}(0) \left( \epsilon_1 {\hat I}_{{\bm 0}}(0) -
\epsilon_2 {\hat I}_{{\bm x}}(0) \right)   +
  {\hat I}_{{\bm r}-{\bm x}}(0) \left( \epsilon_2 {\hat I}_{{\bm 0}}(0) - \epsilon_1 {\hat
I}_{{\bm x}}(0) \right)
  \right\}
 }{1+(\alpha +\beta ){\hat I}_{\bm 0}(0)+\alpha\beta({\hat I}_{\bm 0}^2(0) -
 {\hat I}_{\bm x}^2(0))}.
\end{eqnarray}
\end{widetext}

From now on, for the sake of concreteness, we focus on the case where we have two
competing zealots,
$\epsilon_1=-\epsilon_2=1$, and thus the expression (\ref{Sr3DSTAT}) becomes
$S_{\bm r}(\infty) =\frac{\alpha {\hat I}_{\bm r}(0)- \beta
{\hat I}_{\bm r- \bm x}(0)+\alpha\beta \,
({\hat I}_{\bm r}(0)- {\hat I}_{\bm r-\bm x}(0))({\hat I}_{\bm 0}(0)+
 {\hat I}_{\bm x}(0))}{1+(\alpha +\beta ){\hat I}_{\bm 0}(0)+\alpha\beta({\hat I}_{\bm 0}^2(0) -
 {\hat I}_{\bm x}^2(0))}$.
As we are mainly interested in the large $r$ limit,
one can observe that ${\hat I}_{\bm r}(0)$ is just the static solution of the Poisson
equation
$\Delta_{{\bm r}} {\hat I}_{\bm r}(0)=-\delta_{{\bm r,\bm 0}}$, which solution in the
continuum limit is ${\hat I}_{\bm r}(0)\simeq {\hat {\cal I}}({\bm r})=\frac{1}{4\pi \,
r}$\, ($r> 0$). This result, obtained from an ``electrostatic reformulation'', is valid on
the discrete lattice for $r\gg 1$ \footnote{We have compared the continuum result  to the
exact
expression (\ref{Ir3D}) and have noticed that the discrete and continuum expressions are
very close, even for finite values of $r$: for instance, at site ${\bm r}=(5,3,1)$, we
have exactly ${\hat I}_{\bm r}(0)= 0.01344\dots$, whereas the ``electrostatic''
reformulation gives ${\hat {\cal I}}({\bm r})=\frac{1}{4 \pi \sqrt{35}}=0.01345\dots$
This shows that, already for $r$ finite, the latter reformulation
 is an excellent approximation of (\ref{Ir3D}).}. With the help of (\ref{Sr3DSTAT}), this
result
allows to compute the $3D$ stationary local magnetization for $r\gg  1$ and $|{\bm r} -
{\bm x}| \gg 1$ :
\begin{eqnarray}
\label{Sr3Dstat}
S_{\bm r}(\infty)  = -\frac{1}{4\pi} \left[ \frac{{\cal C}_1}{r} + \frac{{\cal C}_2}{|{\bm
r-\bm x}|} \right],
\end{eqnarray}
where  ${\cal C}_1=-\frac{2\alpha}{2+\alpha g_0}$ and ${\cal C}_2=\frac{2\beta}{2+\beta
g_0}$. Again, the resemblance with electrostatics is striking: the static magnetization is
formally the electrostatic potential generated by the ``charges'' ${\cal C}_1$ at site
${\bm 0}$ and ${\cal C}_2$ at ${\bm x}$. As already noticed, the difficulty resides in the
fact that the charges ${\cal C}_1$ and ${\cal C}_2$
 are {\it a priori} unknown and have been computed in a {\it self-consistent} way (assuming a large enough separating distance $x$), with
the help of the exact and discrete results (\ref{Ir3D}), (\ref{Sr3DSTAT})
 \footnote{In three dimensions, in the presence of $n$ zealots at sites $\{{\bm a}^1,
\dots, {\bm a}^n\}$, using a continuum electrostatic reformulation (which is valid
 if $|{\bm r}-{\bm a}^1| \gg 1, \dots, |{\bm r}-{\bm a}^n|\gg 1$) we
 can infer in the same manner: $S_{\bm r}(\infty)  = -\frac{1}{4\pi} \left[
  \frac{ {\cal C}_1}{|{\bm r}-{\bm a}^1|}+\dots
  + \frac{{\cal C}_n}{|{\bm r}-{\bm a}^n|} \right]$, which can  be developed in multipolar
expansion. In general,  to compute the
  ``charges'' ${\cal C}_1, \dots, {\cal C}_n$ one needs to explicitly invert the
matrix ${\cal N}$. }.
Even though the result (\ref{Sr3Dstat}) is formally valid for $r\gg x\gg 1$, as explained
above,
 it gives already accurate
predictions when $r\gg 1$ and $x$ is finite but large enough ({\it e.g.} already when
$x\geq 6$).
It is suggestive that in the limit where $\alpha=\beta=\infty$, the ``charges''
${\cal C}_{2}=-{\cal C}_{1}=2/g_0$. In this case the magnetization in Eq. (\ref{Sr3Dstat}) can be viewed as
 the potential of the electric dipole of charges $\pm 2/g_0$.
To make the connection with electrostatics even more transparent, it is worthwhile to
notice that the expression (\ref{Sr3Dstat}) can be rewritten using a {\it multipole
expansion}.
Also, when $\beta=0$, we recover $S_{\bm r}(\infty)\propto 1/r$, as reported in Ref.
\cite{IVM}. 
In fact, one has $|{\bm r-\bm x}|^{-1}=(r^2 +x^2 -2{\bm r \bm. \bm
x})^{-1/2}=\frac{1}{r}\sum_{m=0}^{\infty}\left(\frac{x}{r}\right)^{m} \,
P_m(\cos{\theta})$, where  $\cos{\theta} \equiv
\frac{{\bm x \bm. \bm r}}{xr}$ and the $P_m(\cos{\theta})$ are the Legendre polynomials.
Thus the expression (\ref{Sr3Dstat}) can be recast into
\begin{eqnarray}
\label{Sr3Dst}
S_{\bm r}(\infty) =  -\frac{1}{4\pi r}\left[{\cal C}_1  + {\cal C}_2
\sum_{m=0}^{\infty}
\left(\frac{x}{r}\right)^{m} P_m(\cos{\theta})
\right].
\end{eqnarray}
At this point it is important to mention a major difference with the case where only a
single zealot is present. In the latter situation, as showed in Ref.\cite{IVM}, just by
taking the continuum limit of the equation for the magnetization, one could anticipate
that $S_{{\bm r}}(\infty)\propto r^{-1}$ ({\it i.e} it has only radial dependence) in
three dimensions, which is the main desired information. In the two-zealot case, as there
is a competition between the effects of the ``charges'' ${\cal C}_1$ and  ${\cal C}_2$, we
really need to determine $S_{{\bm r}}(\infty)$ through Eqs. (\ref{Ir3D}),(\ref{Sr3DSTAT}),
to  obtain the nontrivial spatial dependence of the stationary magnetization through
(\ref{Sr3Dstat}), (\ref{Sr3Dst}).

Regarding the dynamical approach toward the steady state, it is difficult to study the
small $s$ behavior of ${\hat I}_{\bm r}(s)$ and to  rigorously obtain the long-time
approach toward the stationary magnetization.
However, it follows from Eq. (\ref{formal}) that:

\begin{eqnarray}
\label{Sr3D}
S_{\bm r}(t)-S_{\bm r}(\infty)  &\approx& \frac{1}{2\pi \, (4\pi
t)^{\frac{1}{2}}} \nonumber\\
&\times&\left[ {\cal C}_1 e^{-r^2/4t} +{\cal C}_2\, e^{-|{\bm r-\bm x}|^2/4t}
\right].
\end{eqnarray}
This result is expected to be accurate in the regime where $t\to \infty, r\gg 1$
 and $|{\bm r-\bm x}|\gg 1 $.
As previously mentioned, we can can discuss about the regions with positive or negative
stationary
magnetization. To determine the ``neutral'' region (where $S_{\bm r}(\infty)=0$)
it follows from (\ref{Sr3Dstat}) that, in the limit where $r \gg 1$ and $|{\bm r}-{\bm
x}|\gg 1$,  one has to solve
\begin{eqnarray}
\label{reg0magn3D}
\frac{r}{|{\bm r- \bm x}|}=\delta^{-1} \quad \text{with} \quad
\delta \equiv \left|\frac{{\cal C}_2}{{\cal C}_1}\right| =\frac{\beta(2+\alpha
g_0)}{\alpha(2+\beta g_0)}.
\end{eqnarray}
When $\alpha \neq \beta$, i.e. for $\delta\neq 1$,
 the equation can again be recast into the following form:
$r^2+\frac{2rx}{\delta^2 -1}\, \cos{\theta} - \frac{x^2}{\delta^2 -1}=0$. Such an
expression is the  polar equation of a sphere $\Sigma({\bm C}, {\cal R})$ centered at
${\bm C}=\frac{1}{1-\delta^2}{\bm x}$ with a radius ${\cal R}=\frac{\delta
x}{|1-\delta^2|}$.
From Eq.(\ref{Sr3Dstat}), we can also infer the following:

$\bullet$ If $\delta>1$, {\it i.e.} $\beta>\alpha$, the agents that are within (outside)
the sphere
$\Sigma({\bm c}, R)$ have a positive (negative) magnetization.

$\bullet$ If $\delta<1$, {\it i.e.} $\beta<\alpha$, the
agents that are within (outside) the sphere
$\Sigma ({\bm C}, {\cal R})$ have a negative (positive) magnetization.

These results show that majority of the voters, except the ones enclosed in the 
sphere $\Sigma({\bm C}, {\cal R})$, tends to follow the opinion favored by the strongest 
zealot. The details of the neutral region $\Sigma({\bm C}, {\cal R})$ depend
nontrivially 
on all the parameters $\alpha, \beta$ and ${\bm x}$. In particular, we notice that ${\cal
R}$ 
increases linearly with the separating distance $x$.

$\bullet$ The case where $\alpha=\beta$, {\it i.e.} $\delta=1$, is
special because thus the ``effective charges'' are such that $|{\cal C}_1|={\cal C}_2$.
In particular, this is the case when $\alpha=\beta=\infty$.
It thus follows from (\ref{reg0magn3D}) that the region with zero-final
magnetization is no longer a surface but an infinite plane, given by
$r=\frac{x}{2\cos{\theta}}$, that separates the $3D$ space into two regions.

In 3D, $\alpha(1-S_{\bm 0}(\infty))\neq \beta(1+S_{\bm x}(\infty))$ when $\alpha \neq \beta$, and thus 
the global magnetization of the above inhomogeneous voter model evolves linearly with the time: $M(t)\sim t$. 
This implies that the time $T$ necessary for the strongest
zealot to dominate (on average) the whole 3D system scales as  $T \sim  L^{3}$, where $L\to \infty$.
On the other hand, when $\alpha = \beta$, the space is divided in two symmetric regions with opposite total magnetization.

Finally, in the case where both zealots favor the same opinion $\epsilon=\pm 1$, {\it i.e.} $\epsilon_1=\epsilon_2=\epsilon$, one has just to modify the expressions of ``charges'' in Eqs (\ref{Sr3Dstat}), (\ref{Sr3Dst}) and (\ref{Sr3D}). In fact, these results are still valid with ${\cal C}_1 = -\frac{2\epsilon \alpha}{2+\alpha g_0} $ and ${\cal C}_2 = -\frac{2\epsilon \beta}{2+\beta g_0} $. 

\section{Monomer-monomer catalytic reaction on an inhomogeneous substrate}
The other model that we specifically consider in this work is the monomer-monomer
catalytic
reaction. Such a process is of considerable interest in many fields of science
and the technology. In the catalysis the rate of a chemical reaction is
enhanced by the presence of a suitable catalytic material, such as the platinum
 used to catalyze the oxidation of carbon monoxide ($2CO+O_2\to
2CO_2$) \cite{Boudart,Campbell}. Because of the numerous and practical implications of the
catalytic reaction, it is of prime
interest to be able to model its quantitative and qualitative behavior. In
general, these processes are described within mean-field
like approaches where it is assumed that molecules are randomly distributed on the
substrate \cite{Boudart,Campbell}. Spatial fluctuations
and excluded volume constraints are thus ignored, despite of the fact that these effects
are
 shown to play often a crucial r\^ole \cite{fluct}.

In the modeling of catalysis \cite{Campbell}, the monomer-monomer
surface reaction model plays an important part at least from a theoretical point of
 view because the simplicity of the model allows to address several issues
 analytically, such as the r\^ole
 of the fluctuations \cite{fluct,K1}, the interfacial roughening \cite{roughen},
 and the diffusion of the adsorbents \cite{dif}.

The monomer-monomer catalytic process on an homogeneous substrate is by now well
understood and  it comprises the following reactions \cite{K1,Frachebourg}:
\begin{eqnarray*}
\label{cat}
A+\emptyset &\xrightarrow[k_A]{}&A_{S} \nonumber\\
B+\emptyset &\xrightarrow[k_B]{}& B_{S}\\
A_{S}+B_{S}&\xrightarrow[k_r]{}& AB \,\uparrow + 2\emptyset .
\nonumber
\end{eqnarray*}
The $A$ and $B$ particles impinge upon a substrate with rates $k_A$ and $k_B$,
respectively, adsorb onto vacant sites ($\emptyset$) and form a monolayers of
adsorbed particles, $A_{S}$ and $B_{S}$. Nearest-neighbor pairs
of different adsorbed particles, $A_{S}B_{S}$, react and desorb
with rate $k_r$, leaving two vacant sites ($2\emptyset$) on the substrate.
The dynamics on a spatially homogeneous substrate is most interesting in dimensions $d\leq
2$, when $k_A=k_B$ (otherwise the species with the bigger rate will rapidly
saturate the substrate). In this case
there is coarsening on the substrate induced by fluctuations  and islands of $A_S$
and $B_S$ particles grow. As in Refs. \cite{K1,Frachebourg}, we will consider the
{\it reaction-controlled} limit , where $k_r\ll k_A=k_B$. This limit
turns out to be useful from a technical point of view and, most importantly,
provides qualitatively the same kind of behavior as the general case
\cite{fluct,K1,Frachebourg}.
In the reaction-controlled limit, the substrate quickly becomes fully occupied and
stays covered with $A_S$'s and $B_S$'s for ever (vacancies are immediately refilled).
The kinetics of monomer-monomer substrate reaction model is therefore a {\it
two-state} system that can be mapped onto the voter model
supplemented  by an infinite-temperature Kawasaki exchange process
\cite{K1,Frachebourg}. In fact, in the monomer-monomer catalytic reaction under
consideration, $A_{S}$ and $B_{S}$ desorb and the resulting
empty sites are instantaneously refilled either by $A_{S}
B_{S}$ (no reaction), $A_{S}A_{S}, \, B_{S}B_{S}$ (voter dynamics), or
by  $B_{S}A_{S}$ (Kawasaki exchange
dynamics at infinite temperature).

Clearly, more realistic situations should include the presence of inhomogeneities
which could deeply affect the properties of the system. In fact, real substrates (in $1D$
and $2D$)
display generally some degrees of spatial heterogeneity which are attributed to
imperfections, such as dislocations and defects \cite{InhCat} that modify locally the
interactions on the substrate. In some previous works translationally-invariant
disordered models for catalysis have been considered within mean-field like approaches, {\it i.e.} rate 
equations and pair approximation \cite{disorder}. In these works, it was shown that
quenched substrate imperfections dramatically affect the dynamics  resulting in a
reactive steady-state. One should emphasize that both the physical systems (in this work, the inhomogeneities are not randomly distributed but fixed) and the analytic methods (we obtain exact results in arbitrary dimensions, while the authors of \cite{disorder} employed mean-field-like approaches) considered here differ from, and are thus complementary to, those of Ref.\cite{disorder}. 
Also, very recently, an {\it equilibrium} model for monomer-monomer
catalysis on a disordered substrate was solved \cite{Oshanin}.

Hereafter we study the static and dynamical  effects of local inhomogeneities in the
monomer-monomer catalytic reaction-controlled process and show how to take advantage of the results
obtained for the inhomogeneous voter model to  infer some exact properties.
In fact, we consider the {\it genuine nonequilibrium} situation where the substrate is
{\it spatially inhomogeneous}, because of the presence of a collection of
$n$ inhomogeneities located at sites $\{{\bm a}^j\}, \, j=1,\dots, n$ favoring
the local adsorption of $A$'s or $B$'s. We show that the inhomogeneities induce
spatially dependent reactive steady-state when $n>1$. As a substrate, as described in Section II, we consider 
an hypercubic lattice with $(2L+1)^d$ sites and
introduce a set of parameters $\epsilon'_j$ taking the values $0$
or $1$ and consider, in addition to the usual homogeneous catalytic
reaction described above, that some inhomogeneities {\it locally} favor
the presence of $A$ via desorption of $B$'s (and vice versa) through
the additional reactions $B_{S}\xrightarrow[\alpha_j]
{}A_{S}$, where $\epsilon'_j=1$, and
 $A_{S}\xrightarrow[\alpha_{j'}]{}B_{S}$, where $\epsilon_{j'}'=0$.
 We therefore consider the following homogeneous
 processes (voter +
 infinite-temperature Kawasaki dynamics), all occurring with the same rates
 $1/2$, and local (inhomogeneous) reactions at
 sites ${\bm a}^j$ and ${\bm a}^{j'\neq j}$, occurring respectively with
 rates $\alpha_j$ and $\alpha_{j'}$:
 \begin{eqnarray*}
 \label{reac}
A_{S}B_{S}&\xrightarrow[1/2]
{}&A_{S}A_{S}; \; A_{S}B_{S}\xrightarrow[1/2]
{}B_{S}B_{S};
\nonumber\\
\; A_{S}B_{S}&\xrightarrow[1/2]
{}&B_{S}A_{S};\; \\
\label{loc}
A_{S}&\xrightarrow[\alpha_j; \,\epsilon'_j=0]
{}&B_{S}; \; B_{S}\xrightarrow[\alpha_{j'}'; \,\epsilon'_j=1]
{}A_{S}.
 \end{eqnarray*}
 Here, the bimolecular reactions correspond to the voter dynamics supplemented by
 Kawasaki infinite-temperature exchange process, whereas
 monomolecular processes correspond to reactions induced by local
 inhomogeneities favoring the adsorption of one species.
 Following the same steps as in Refs \cite{K1,Frachebourg}, for this spatially
 inhomogeneous monomer-monomer catalytic process, in the thermodynamic limit we obtain the following
 equation of motion for the concentration $c_{\bm r}(t)$ of $A_{S}$
  at site ${\bm
 r}$ of the substrate:
 \begin{eqnarray}
 \label{concA}
 \frac{d}{dt}c_{\bm r}(t) = \Delta_{\bm r}c_{\bm r}(t)+\sum_{j=1}^{n} \alpha_j \,
(\epsilon'_j -
 c_{{\bm a}^j}(t))\delta_{{\bm r},{\bm a}^{j}}.
 \end{eqnarray}
Of course, the concentration of $B_S$ at site
${\bm r}$ is simply
given by $1-c_{\bm r}(t)$. The resemblance of Equation (\ref{concA}) with
(\ref{MS}) is striking (the only difference is that here $\epsilon'_j=0,1$)
 and one can immediately infer the solution of (\ref{concA}) from
  (\ref{Sr} and (\ref{Srcont})). In particular, in the thermodynamic limit,
  starting from a system initially completely occupied by $B_S$ particles, the
  Laplace transform of the concentration of $A_s$ reads:
\begin{eqnarray}
\label{concLapl}
{\hat c}_{\bm r}(s)=\frac{1}{s}\sum_{j,\ell}\epsilon'_{\ell} {\hat I}_{{\bm a}^j -{\bm r}}
(s)[{\cal N}^{-1} (s,\{\alpha\})]_{j,\ell},
\end{eqnarray}
and we get for the time-dependent concentration (initially $c_{\bm r}(0)=0$):
\begin{eqnarray}
\label{conc}
c_{\bm r}(t)=\frac{1}{2\pi i}\int_{c-i\infty}^{c+i\infty}\frac{ds}{s}\, e^{st}
\sum_{j,\ell}\epsilon'_{\ell} {\hat I}_{{\bm a}^j -{\bm r}}
(s)[{\cal N}^{-1}]_{j,\ell}.
\end{eqnarray}
%
In this language, the quantity
\begin{eqnarray}
\label{globalA}
M'(t)\equiv \sum_{{\bm k}}c_{{\bm k}}(t)=\sum_{j=1}^{n}\, \alpha_j \int_{0}^{t}d\tau \,
\left[
\epsilon'_j -c_{{\bm a}^{j}}(\tau)
\right]
\end{eqnarray}
provides the average total number of the $A_S$ particles
on the substrate at time $t$.

Next, we restrain ourself to physical situations and consider in detail
the monomer-monomer catalytic reaction in the presence
of one and two inhomogeneities in one and two dimensions.

\subsection{Inhomogeneous monomer-monomer catalytic reaction in the presence
of one single ``defect''}
Here, we consider the case where there is a single
 inhomogeneity  at site ${\bm a}^{1}={\bm 0}$,
 with strength $\alpha_1=\alpha$ and
$\epsilon_1=1$. In this case, we simply have ${\cal N}^{-1}=\frac{\alpha}
{1+\alpha{\hat I}_{0}(s)}$. Therefore, starting from a system initially
full of $B_S$ particle ({\it i.e.}  $c_r(0)=0$) we obtain:
\begin{eqnarray}
\label{cs1inh}
{\hat c}_{r}(s)=\frac{1}{s} \, \frac{\alpha {\hat I}_r}{1+\alpha {\hat I}_0
(s)}.
\end{eqnarray}
On the right-hand side of this equation, one recognizes immediately
the same expression as the Laplace transform of the magnetization
obtained in Ref. \cite{IVM}.
From previous results, we can immediately infer the long-time behavior of
the concentration of $A_s$ particles.

\subsubsection{Results in 1D}

Following the same steps as in Ref. \cite{IVM}, on a one-dimensional
substrate we find from (\ref{cs1inh}) that the long-time behavior of the
concentration of $A_S$ reads:
\begin{eqnarray}
\label{cr1d1inh}
c_r(t)\simeq 1-\frac{r+2/\alpha}{\sqrt{\pi t}}.
\end{eqnarray}
This result is valid for any $0\leq r<\infty$.

When both $r\to \infty$ and $t\to \infty$, we
obtain the following simple
scaling expression \cite{IVM}:
\begin{eqnarray}
\label{cr1d1inhprim}
c_r(t)\simeq {\rm erfc}\left(\frac{r}{2\sqrt{t}}\right).
\end{eqnarray}

\subsubsection{Results in 2D}

In two dimensions,  following results from Ref.\cite{IVM} we obtain a non-scaling
expression for the concentration, with very slow time-dependence:
\begin{eqnarray}
\label{c02D}
c_{\bm 0}(t) -  c_{\bf 0}(\infty)\simeq -
\left(\frac{4\pi }{\alpha}\right)\frac{1}{\ln{t}},
\end{eqnarray}
where $c_{\bm 0}(\infty)=1$. For the other sites,
we find that the long-time behavior in the regime
$t\gg r^{2}\gg 1$ is given by
\begin{eqnarray}
\label{cr2D}
c_{\bm r}(t) -
c_{\bm r}(\infty) \simeq -
\frac{\ln{r^{2}}}{\ln{t}}, \quad c_{\bm r}(\infty) =1.
\end{eqnarray}
As in the one-dimensional case, the stationary concentration of $A_S$
corresponds again to a substrate fully covered with $A_S$ particles, {\it i.e.}
$ c_{\bm r}(\infty) = 1$.
Therefore, the presence of a single
inhomogeneity favoring locally the adsorption of $A_S$  is enough to completely
cover the substrate with $A_S$ in spite of the fact
that initially only $B_S$ particles were present. From the expressions
(\ref{cr1d1inh}), (\ref{c02D}) and (\ref{globalA}), we can also compute the total
number of $A_S$ particles on the substrate at time $t\gg 1$. In so doing, one obtains $M'(t)\sim \sqrt{t}$
 in the one-dimensional case and $M'(t)\sim t/\ln{t}$ in 2D.

\subsection{Inhomogeneous monomer-monomer catalytic reaction in the presence
of two defects}
Here, we consider the case where two ``competing'' inhomogeneities are present: one is
 at site ${\bm a}^{1}={\bm 0}$, with strength $\alpha_1=\alpha$ and
$\epsilon_1=1$ and the other at site ${\bm a}^{2}={\bm x}$, with strength
 $\alpha_2=\beta$ and $\epsilon_2=0$.

In this case, using Eqs.(\ref{concLapl}) and (\ref{N}), we obtain the following
expression for the Laplace transform of the concentration of $A_s$ at site
${\bm r}$, starting from $c_{\bm r}(0)=0$:
\begin{eqnarray}
\label{crcont2inh}
&&\hat{c}_{\bm r}(s) =\frac{1}{s}\sum_{j,\ell}\,
 {\hat I}_{{\bm a}^j -{\bm r}}(s) \,\epsilon'_\ell [{\cal N}^{-1} (s,\{\alpha\})]_{j,\ell} \nonumber\\
&=& \frac{\alpha {\hat I}_{\bm r}(s)+\alpha\beta \,
({\hat I}_{\bm r}(s) {\hat I}_{\bm 0}(s) -
{\hat I}_{\bm r-\bm x}(s){\hat I}_{\bm x}(s))}{s\, [1+(\alpha +\beta ){\hat I}_{\bm 0}(s)+
\alpha\beta({\hat I}_{\bm 0}^2(s) - {\hat I}_{\bm x}^2(s))]}.
\end{eqnarray}
\subsubsection{Results in 1D}

In one dimension, without loss of generality, we assume
that the inhomogeneity at site ${\bm a}^2={\bm x}$, $x=|{\bm x}|$,  is on the
right side of the origin.

Proceeding as in section IV.A, we study the static and long-time behavior
 of the concentration of $A_S$ with $c_r(0)=0$, and distinguish various situations:
\begin{itemize}
\item For sites between the two inhomogeneities, {\it i.e.}  $0\leq r\leq x$ we get:
\begin{eqnarray}
\label{crbetween}
&& c_{r}(t)\simeq \frac{\alpha[1+\beta(x-r)]}{\alpha
+\beta+\alpha\beta x} 
- \left(\frac{\alpha}{\alpha
+\beta+\alpha\beta x} \right) \nonumber\\ &\times&
\frac{1}{\sqrt{\pi t}}\left\{r+\frac{[1+\beta(x-r)](2-\alpha\beta x^2/2)}
{\alpha+\beta+\alpha\beta x}\right\}
\end{eqnarray}
\item At the right of the origin, when $x<r<\infty$, we obtain:
\begin{eqnarray}
\label{crright}
c_{r}(t)&\simeq& \frac{\alpha}{\alpha
+\beta+\alpha\beta x} \nonumber\\ &\times&\left[
1-\frac{1}{\sqrt{\pi t}}\,\left\{
r+\frac{2-\alpha\beta x^2/2}{\alpha+\beta+\alpha\beta x}
\right\}
\right].
\end{eqnarray}
\item At the left of the origin, when $0<r<\infty$, we find:
\begin{eqnarray}
\label{crleft}
c_{-r}(t)&\simeq& \frac{\alpha}{\alpha
+\beta+\alpha\beta x} \nonumber\\ &\times&\left[
1+\beta x-\frac{1}{\sqrt{\pi t}}\,\left\{
r+ \frac{(1+\beta x)(2-\alpha\beta x^2/2)}{\alpha+\beta+\alpha\beta x}
\right\}
\right].\nonumber\\
\end{eqnarray}
\item When both $t\to \infty$ and $r\to \infty$, we have:
\begin{eqnarray}
\label{crscal1}
c_{r}(t) &\simeq &\frac{\alpha}{\alpha
+\beta+\alpha\beta x}\, {\rm erfc}\left(\frac{r}{2\sqrt{t}}\right) \\
\label{crscal2}
c_{-r}(t) &\simeq & \frac{\alpha (1+\beta x)}{\alpha
+\beta+\alpha\beta x}\, {\rm erfc}\left(\frac{r}{2\sqrt{t}}\right).
\end{eqnarray}
\end{itemize}
These results show that in the interval between the of inhomogeneities, the static
concentration
profiles varies linearly from the origin with a slope
$-\alpha\beta/(\alpha+\beta+\alpha\beta x)$.
Outside from this interval, the static concentration is uniform on the right and left side
of
the origin: on the right, $c_{r}(\infty)=\frac{\alpha (1+\beta x)}{\alpha
+\beta+\alpha\beta x}$, whereas on the left $c_{r}(\infty)=\frac{\alpha }{\alpha
+\beta+\alpha\beta x}$.
Such a static profile can again be
interpreted as the solution of a discrete  1D electrostatic
Poisson equation with peculiar and suitable boundary conditions.
Again, the static concentration is reached according to a power-law ($c_r(t) \sim
t^{-1/2}$) and with amplitudes depending nontrivially on all parameters of the
system. At very large distances, and long time, the concentration displays
a scaling form which amplitude depends on which inhomogeneity is the
closest.
Of course, it is easy to check that in the limit $\alpha\to 0$ , as the system is
initially full of $B_S$, then $c_r(t)=0$. Also, when $\beta=0$,
we recover the  expressions (\ref{cr1d1inh}) and (\ref{cr1d1inhprim}).
From Eqs (\ref{globalA}) and (\ref{crbetween}) we obtain the average number of adsorbed particles which
 evolves (at long-time) as $M'(t)\sim \sqrt{t}$.

Again, in one dimension we can obtain the stationary concentration of adsorbed $A_S$
particle in the completely disordered case, {\it i.e} when $n$ is arbitrary large just by
replacing respectively $S_{a^j}(\infty), S_r(\infty), \epsilon_j$ by $c_{a^j}(\infty),
c_r(\infty), \epsilon'_j$ in the expressions (\ref{Saj})-(\ref{magnstat_arb3}). As illustrated 
in Fig.\ \ref{1d_4z}, in this case the stationary concentration profile is piecewise. 
Also, when the number of competing inhomogeneities is finite the system coarsens as described in
 Section IV.A.

\subsubsection{Results in 2D}

In two dimensions and at large distance from both inhomogeneities, {\it i.e }
 for $r \gg 1$ and $|{\bm r}-{\bm x}|\gg 1$, we find a non-scaling expression for both
static and
 time-dependent concentration
 of the $A_S$ particles:
\begin{eqnarray}
\label{cr2dstat}
 c_{\bm r}(\infty)
\simeq
\frac{\alpha - \frac{\alpha \beta}{2\pi} \left[\ln{\left(\frac{r}{x|{\bm r}-
{\bm x}|}\right)}-\pi(\gamma -\ln{2}) \right]}{\alpha +\beta +\frac{\alpha
\beta}{\pi}(\ln{x}+\pi(\gamma -\ln{2}))},
\end{eqnarray}
and, when $x$ is large enough, $c_{\bm 0}(\infty)\simeq
\frac{\alpha\left(1+\frac{\beta}{\pi}\ln{x}\right)}{\alpha +\beta+\frac{\alpha \beta}{\pi} \ln{x}}$ and 
$c_{\bm x}(\infty) \simeq \frac{\alpha}{\alpha +\beta+\frac{\alpha \beta}{\pi} \ln{x}}$.

We can notice that in $2D$ the stationary concentration of the $A_S$ particles is a
fluctuating reactive state exhibiting nontrivial radial and polar dependence. Regarding
the approach toward the steady state, proceeding as in the section IV.B, we obtain:
\begin{eqnarray}
\label{cr2d}
c_{\bm r}(t)-c_{\bm r}(\infty) \simeq
-\frac{B'({\bm r},{\bm x})}{\ln{t}}
\end{eqnarray}
where the amplitude $B'= \frac{\frac{\alpha \beta}{\pi} \ln{x}\left\{
\ln{|{\bm r}-{\bm x}|} +\gamma - \ln{2}
\right\} +2\alpha \ln{r} }{\alpha +\beta +\frac{\alpha \beta}{\pi}
[\ln{x}+\pi(\gamma -\ln{2})]}$ exhibits a nontrivial spatial dependence.
Again, the result (\ref{cr2d}) shows that the  stationary concentration profile
(\ref{cr2dstat})
is reached logarithmically slowly. 
Using Eq. (\ref{globalA})  we can also notice that the average number of particles $A_S$ adsorbed on the substrate evolves (at long-time) as
$M'(t)\sim t/\ln{t}$.

There is a practical interest in understanding the spatial distribution of adsorbed
particles
in the steady state \cite{exp} and one can thus ask:{\it What is the region  of
the 2D substrate where one can find more $A_S$ particles ?}

To answer this question, from Eq. (\ref{cr2dstat}), we proceed as in Section IV.B and,
according to Eq.(\ref{reg0magn}), we see that when $\alpha > \beta$  ($\beta > \alpha$),
the region richer in $A_S$ particles is outside (within) the disk
{\rm Int}$\,{\cal C}({\bm c}, R)$ [defined in Section IV.B], where the concentration of $A_S$ is $c_{\bm
r}(\infty)\geq\frac{1}{2}$
($c_{\bm r}(\infty)\leq\frac{1}{2}$). When $\alpha=\beta$, the 2D substrate is separated
into two half-planes with concentration of $A_S > 1/2$ in the region including the origin.

\section{Summary and Conclusion}

In this work we have shown how to compute some exact properties of a class of
many-body stochastic systems in the presence of an arbitrary number  of
inhomogeneities $n$, and have specifically focused on the voter model and
monomer-monomer catalytic reaction (in the reaction-controlled limit).
We have studied the effects of local perturbations of the dynamical rules on the static
and
time-dependent properties of these models
by obtaining both general (yet formal) and many explicit results in the presence of one
and two inhomogeneities.
In fact, the latter situation already displays and covers most of the generic features of the
models.
 Namely, when there is only one inhomogeneity present, it is responsible for a uniform and
``unanimous'' steady state in low dimensions \cite{IVM}, while in the presence of
competing
inhomogeneities ($n>1$) the steady state is fluctuating and reactive.
For the sake of concreteness we have mainly focused on the amenable case with two
inhomogeneities and have shown quantitatively how the local interactions 
deeply affect the properties of these systems.
Neither the stationary nor the time-dependent expression of the order parameters
are translationally-invariant but exhibit nontrivial radial and polar dependence (when
$d>1$).

From a sociophysical perspective, in the voter model language, this means that a system
which tolerates the presence of ``competing zealots'', {\it i.e.} which accepts the
competition between opposite points of view, will never reach a unanimous state but 
always end into a final configuration where both opinions coexist and fluctuate.
Of course, such a conclusion seems to be consistent with the results of electoral
competitions in modern democracies.

In the presence of competing inhomogeneities ($n>1$) in low dimensions, subtle 
coarsening phenomena take place in 1D and 2D. In fact, the local and competing perturbations of 
the dynamics lead us to distinguish the case where the number of inhomogeneities is finite
and the case where their number is comparable to the size of the system. In the former case the system 
coarsens and large domains develop, but their size are typically limited by the number of competing 
inhomogeneities, while in the latter case coarsening is prevented by the interaction with all the 
numerous inhomogeneities.

More specifically, in this work we have obtained exact, yet formal,
expressions of the static and time-dependent order parameters (see (\ref{Sr}) and (\ref{conc})).
The main technical problem to carry out
detailed calculation resides in the inversion of the $n \times n$ matrix
${\cal N}$. The case with one single inhomogeneity in the voter model was already
considered in \cite{IVM} and here we show that such results can be
translated in the language of the catalysis reaction. In particular we have
shown that on 1D and 2D substrates, the presence of a single spatial
inhomogeneity favoring the adsorption of one species, say $A_S$, with respect
to the other is sufficient to ensure that eventually the substrate
will be completely filled with $A_S$ particles.
When we have two competing inhomogeneities, favoring locally opposite states or the
adsorption of particles of different species, we have obtained rich
behavior. In 1D, between the two inhomogeneities, the stationary profiles of the
order parameters vary linearly with the distance from the origin
(\ref{Sr1Drfin}),(\ref{crbetween}) and then reaches two plateaus (\ref{Sr1Drfinprim}),
(\ref{Sr1Drfinprimprim}) and (\ref{crright}), (\ref{crleft}).
  These static profiles are always reached algebraically in 1D:
  $S_r (t)-S_r(\infty)\simeq At^{-1/2}$ and $c_r (t)-c_r(\infty)\simeq
  A't^{-1/2}$, where the amplitudes $A$ and $A'$ depend nontrivially on all
  parameters of the problem and in particular on the separating distance between the
inhomogeneities [see Eqs (\ref{Sr1Drfinprim}), (\ref{Sr1Drfinprimprim}) and
(\ref{crright}), (\ref{crleft})].
  Far away from the inhomogeneities, the order parameters display scaling
  expression of the variable $r/\sqrt{t}$ [see (\ref{Sr1Dscal}) and (\ref{crscal1}),
   (\ref{crscal2})]. In one dimension, we have also been able to compute the expression of
the stationary magnetization in the completely disordered situation where the number of
zealots is arbitrary large [see Eqs. (\ref{Saj})-(\ref{magnstat_arb3})].
   In two dimensions, for $n=2$, in the presence of two competing inhomogeneities,
   we have obtained non-uniform and nontrivial
   stationary profiles for the order parameters, in agreement with
   an electrostatic-like reformulation, the latter display logarithmic spatial dependence
(radial and polar) [(\ref{Sr2dstat}) and (\ref{cr2dstat})]. The approach toward the
reactive steady state
    is very slow: $S_r (t)-S_r(\infty)\simeq B/\ln{t}$ and
   $c_r (t)-c_r(\infty)\simeq
  B'/\ln{t}$, with amplitudes $B$ and $B'$ depending again nontrivially on all
  parameters of the problem [see (\ref{Sr2ddyn}) and (\ref{cr2d})].
  In 2D, for the inhomogeneous voter model, we have also studied the spatial
  regions with positive/negative static magnetization and have shown that only within
   a circle, whose center and radius depend on the strength of the ``zealots" and on the
  distance between the latter, the sign of the magnetization is the one favorite
  by the ``weakest'' zealot. When both zealots have the same strength, there is
  positive/negative magnetization in half-space.
In three dimensions, for $n=2$ and in the continuum limit, we have shown that the
stationary magnetization of the inhomogeneous voter model displays a radial and polar
dependence that can be recast into a multipole expansion (\ref{Sr3Dstat}), corresponding
formally to the electrostatic potential generated by two ``charges'' that are determined
self-consistently using exact results from the discrete lattice system. The connection with electrostatics is 
particularly striking in the limit where both zealots have an infinite strength, thus the stationary 
magnetization corresponds to the potential of an electric dipole.
The approach toward the static magnetization follows a power-law: $S_{\bm r}(t)-S_{\bm
r}(\infty)\simeq C t^{-1/2}$ (see (\ref{Sr3D})). Also, in 3D we have studied the
spatial regions with positive/negative magnetization and have shown that outside
from a sphere whose center and radius depend on the parameters of the system and varies
linearly with the distance separating the zealots, the sign of the final magnetization is the 
one favored by the strongest zealot.

The results obtained from Monte Carlo simulations of one and two-dimensional lattices
show excellent agreement with the theoretical results obtained for an infinite system.
In the presence of multiple ($n>2$) competing inhomogeneities the calculations in two dimensions 
become very tedious and we consider this case by numerical simulations which confirm the extremely slow dynamics and the existence of nontrivial spatial dependence of the order parameters. 
We also would like to point out one intriguing and interesting fact about the small
time behavior of the magnetization of the zealots in the one-dimensional case. As it can
be extracted from Fig.\ \ref{1d_2z}, $S_0(t)$ and $S_x(t)$, for small $t$, evolve as
a power law with an exponent numerically smaller than $0.50$. The small time behavior
of the site magnetization of the usual one-dimensional voter model 
(no inhomogeneities) is linear, i.e. $S_r(t) -S_{r}(0)\propto t$ for any site $r$ on the lattice. 
We think it would be interesting to investigate further this ``anomalous'' small-$t$ behavior 
of the magnetizations of the zealots in the one and the two-dimensional cases and we plan 
to do it in our future work.
Various generalizations of this work could also be investigated.
For instance, it would be worthwhile to consider that the inhomogeneities
would not be fixed but spatially distributed according to some function 
${\cal P}(\{ {\bm a}^j \})$. In this case, one should also average on the quenched
disorder (on the samples) and one would have to compute: ${\bar S}_{\bm
r}(t)\propto \sum_{\{{\bm a}^j\}} {\cal P}(\{ {\bm a}^j \}) S_{\bm r}(\{ {\bm a}^j \} ,t)$, where
$S_{\bm r}(\{ {\bm a}^j \},t)$ is the quantity studied in this work for a given set of
inhomogeneities at sites 
$\{ {\bm a}^j \}$.
In the same manner, it would be quite interesting to consider the disordered case where
the strength of the inhomogeneities would follow a distribution function such as
${\widetilde {\cal P}}(\{\alpha_j\})\propto \prod_{j=1}^{n} e^{-(\alpha_j -{\bar
\alpha})^2/2\sigma}$. In this case, one would be interested in the quantity:
${\widetilde S}_{\bm r}(t)=\int \prod_{j}d\alpha_j \,
{\widetilde {\cal P}}(\{\alpha_j\}) S_{\bm r}(\{ {\bm a}^j \}, \{
{ \alpha}^j \},t)$, where
$S_{\bm r}(\{ {\bm a}^j \},\{ {\alpha}^j \}, t)$ is the magnetization computed in this
work for a given set of 
inhomogeneities at sites  $\{ {\bm a}^j \}$, with strength $\{ {\alpha}^j \}$. 

\section{Acknowledgments}
We are grateful to B. Schmittmann, U. C. T\"auber and R. K. P. Zia for numerous fruitful
discussions, advices and suggestions. We thank P. L. Krapivsky for bringing Ref.\cite{Muk} to our attention and for useful comments. MM acknowledges the financial support of Swiss NSF
Fellowship No. 81EL-68473. This work was also partially supported by US NSF grants
DMR-0088451, 0308548 and 0414122.

\end{document}